\documentclass[12pt,prd,showpacs,tightenlines,nofootinbib]{revtex4}
\usepackage{bm}
\usepackage{graphics}
\usepackage{rotating}
\usepackage{epsfig}
\usepackage{longtable}
\begin{document}
\begin{flushright}{HU-EP-11/21}\end{flushright}
\title{Spectroscopy and Regge trajectories of heavy baryons in the
  relativistic quark-diquark picture}
\author{D. Ebert$^1$, R. N. Faustov$^{1,2}$  and V. O. Galkin$^{1,2}$}
\affiliation{$^1$ Institut f\"ur Physik, Humboldt--Universit\"at zu Berlin,
Newtonstr. 15, D-12489  Berlin, Germany\\
$^2$ Dorodnicyn Computing Centre, Russian Academy of Sciences,
  Vavilov Str. 40, 119991 Moscow, Russia}

\begin{abstract}
Mass spectra of heavy baryons are calculated in the
heavy-quark--light-diquark picture in the framework of the QCD-motivated
relativistic quark model. The
dynamics of light quarks in the diquark as well as the dynamics of the
heavy quark and light diquark in the baryon are treated completely
relativistically without application of nonrelativistic $v/c$
and heavy quark $1/m_Q$ expansions. Such approach allows us to get
predictions for the heavy baryon masses for rather high orbital
and radial excitations. On this basis the Regge trajectories of heavy baryons for
orbital and radial excitations are constructed, and their linearity, 
parallelism, and equidistance are verified. The relations between the
slopes and intercepts of heavy baryons are considered and a comparison
of the slopes of Regge trajectories for heavy baryons and heavy-light
mesons is performed. All
available experimental data on heavy baryons fit nicely to the
constructed Regge trajectories. The possible assignment of the quantum
numbers to the observed excited charmed baryons is discussed.  
\end{abstract}

\pacs{14.20.Lq, 14.20.Mr, 12.39.Ki}

\maketitle

\section{Introduction}
\label{sec:int}

Recently a significant experimental progress has been achieved in
studying the heavy baryon spectroscopy.  In the last five years the
number of the observed charmed and bottom baryons almost doubled and
now it is nearly the same as the number of known charmed and bottom
mesons \cite{pdg}. Observations of new  charmed baryons were mainly done
at the $B$-factories, while new bottom baryons were discovered at
Tevatron \cite{kr}. It is expected that new data on excited bottom baryons
will come soon from the LHC, where they are supposed to be copiously
produced. Due to the poor statistics, the quantum numbers of most of
the excited states of heavy baryons are 
not known experimentally and are usually
prescribed following the quark model predictions \cite{pdg}.

In this paper we investigate heavy baryon spectroscopy in the
framework of the QCD-motivated relativistic quark model based on the
quasipotential approach \cite{egf,hbar}. To simplify the very complicated relativistic
three-body problem heavy baryons are considered in the
heavy-quark--light-diquark approximation. This reduces the initial three-body
problem to two step two-body calculations. First, the light diquark properties, such
as masses and form factors, are presented \cite{hbar}. Then a heavy baryon is
considered as the bound system of a heavy quark and a light
diquark. In order to take into account the rather large size and structure of the
light diquark, its nonlocal interaction with gluons is described by the
form factor expressed in terms of the diquark wave functions. All
heavy baryon excitations, both orbital 
and radial, are assumed to occur in the bound system of the heavy quark and light
diquark, while the latter is taken only in the ground (scalar or axial
vector) state. Such scheme significantly reduces the number of the
excited baryon states compared to the genuine three-quark picture.        
The goal of this paper is the calculation of the masses of the excited
heavy baryons up to rather high orbital and radial excitations. This
will allow us to construct the heavy baryon Regge trajectories both in
the $(J,M^2)$ and $(n_r,M^2)$ planes, where $J$ is the baryon
spin, $M$ is the baryon mass and $n_r$ is the radial quantum
number. Then we can test their linearity, parallelism and equidistance
and determine their parameters: Regge slopes and intercepts. Their
determination is of great importance, since they provide a better
understanding of the hadron dynamics. Moreover, their knowledge is also
important for non-spectroscopic problems such as, e.g., hadron
production and high energy scattering.  
Since we are going to calculate highly excited heavy baryon states it
is important to use a fully relativistic approach, which does not use the
nonrelativistic $v/c$ expansion for light quarks and diquarks  and
does not employ the heavy quark $1/m_Q$ expansion for the heavy quark.

The heavy baryon spectroscopy has been extensively studied in the
literature
\cite{kr,ci,mmmp,gvv,qm,jen,lat,ww,sr,hbar,excbar}. Various quark
models \cite{kr,ci,mmmp,gvv,qm,hbar,excbar}, heavy quark $1/m_Q$ and
$1/N_c$ expansions \cite{jen}, quenched and unquenched lattice
calculations \cite{lat,ww} and QCD sum rules \cite{sr} have been
used. However, in all these calculations either masses of the ground
state baryons were obtained or only a few lowest orbital and radial
excitations were considered. Therefore the Regge trajectories of heavy
baryons have not been constructed. Contrarily, the Regge trajectories
of light baryons received significant attention
\cite{kr,m,s,bg,is,aamnsv,cl,f,bt}. The related investigations were performed on the
basis of quark models \cite{m,s,bg,is}, empirical relations
\cite{cl}  and in models based on the AdS/QCD duality
\cite{f,bt}. It was shown that the highly orbitally excited light 
baryons have an antisymmetric structure of the quark-diquark type \cite{m,s} and
such configuration minimizes the energy \cite{m}. Only
in this case light baryon and meson Regge trajectories have the same
slope \cite{m,s}~\footnote{Note
that all these considerations were done for massless scalar quarks.} which is in agreement with experimental data.  

Several simple relations between slopes and intercepts of light and
heavy baryons have been deduced in different models within QCD (see,
e.g., \cite{bg,k,gww} and references therein). They were used for
obtaining various linear and quadratic mass relations between baryon
masses \cite{gww}. 

The paper is organized as follows. In Sec.~\ref{sec:rqdm} we present
the relativistic quark-diquark model of heavy baryons. First we
discuss properties of light diquarks and give their masses and form
factors. Then a heavy baryon is considered as the bound system of a heavy
quark and a light diquark. The completely relativistic expressions for
the corresponding quasipotentials are given. In Sec.~\ref{sec:rd}
the heavy baryon spectroscopy is presented and discussed. 
Our predictions for charmed and bottom baryon masses are confronted 
with the available experimental data. The obtained results are used for
constructing the heavy baryon Regge trajectories both in
the $(J,M^2)$ and $(n_r,M^2)$ planes. The prescription of the
observed baryon states to the particular trajectory allows
to determine their quantum numbers. Then we obtain slopes and
intercepts of parent and daughter trajectories and test the proposed relations
between them.  Finally, a comparison of the slopes of the heavy meson and
heavy baryon Regge trajectories is performed. We present our conclusions
in Sec.~\ref{sec:conc}.

\section{Relativistic quark-diquark model of heavy baryons}
\label{sec:rqdm}

In the quasipotential approach and quark-diquark picture of
heavy baryons the interaction of two light quarks in a diquark and the heavy
quark interaction with a light diquark in a baryon are described by the
diquark wave function ($\Psi_{d}$) of the bound quark-quark state
and by the baryon wave function ($\Psi_{B}$) of the bound quark-diquark
state respectively,  which satisfy the
quasipotential equation of the Schr\"odinger type \cite{egf}
\begin{equation}
\label{quas}
{\left(\frac{b^2(M)}{2\mu_{R}}-\frac{{\bf
p}^2}{2\mu_{R}}\right)\Psi_{d,B}({\bf p})} =\int\frac{d^3 q}{(2\pi)^3}
 V({\bf p,q};M)\Psi_{d,B}({\bf q}),
\end{equation}
where the relativistic reduced mass is
\begin{equation}
\mu_{R}=\frac{M^4-(m^2_1-m^2_2)^2}{4M^3},
\end{equation}
and $E_1$, $E_2$ are given by
\begin{equation}
\label{ee}
E_1=\frac{M^2-m_2^2+m_1^2}{2M}, \quad E_2=\frac{M^2-m_1^2+m_2^2}{2M}.
\end{equation}
Here $M=E_1+E_2$ is the bound state mass (diquark or baryon),
$m_{1,2}$ are the masses of light quarks ($q_1$ and $q_2$) which form
the diquark or of the light diquark ($d$) and heavy quark ($Q$) which form
the heavy baryon ($B$), and ${\bf p}$  is their relative momentum.  
In the center of mass system the relative momentum squared on mass shell 
reads
\begin{equation}
{b^2(M) }
=\frac{[M^2-(m_1+m_2)^2][M^2-(m_1-m_2)^2]}{4M^2}.
\end{equation}

The kernel 
$V({\bf p,q};M)$ in Eq.~(\ref{quas}) is the quasipotential operator of
the quark-quark or quark-diquark interaction. It is constructed with
the help of the
off-mass-shell scattering amplitude, projected onto the positive
energy states. In the following analysis we closely follow the
similar construction of the quark-antiquark interaction in mesons
which were extensively studied in our relativistic quark model
\cite{egf}. For
the quark-quark interaction in a diquark we use the relation
$V_{qq}=V_{q\bar q}/2$ arising under the assumption about the octet
structure of the interaction  from the difference of the $qq$ and
$q\bar q$  colour antitriplet and singlet states. An important role in this construction is
played by the Lorentz-structure of the nonperturbative confining  interaction. 
In our analysis of mesons, while  
constructing the quasipotential of the quark-antiquark interaction, 
we adopted that the effective
interaction is the sum of the usual one-gluon exchange term with the mixture
of long-range vector and scalar linear confining potentials, where
the vector confining potential contains the Pauli term.  
We use the same conventions for the construction of the quark-quark
and quark-diquark interactions in the baryon. The
quasipotential  is then defined by the following expressions \cite{efgm,egf} 

(a) for the quark-quark ($qq$) interaction in the colour antitriplet state
 \begin{equation}
\label{qpot}
V({\bf p,q};M)=\bar{u}_{1}(p)\bar{u}_{2}(-p){\cal V}({\bf p}, {\bf
q};M)u_{1}(q)u_{2}(-q),
\end{equation}
with
\[
{\cal V}({\bf p,q};M)=\frac12\left[\frac43\alpha_sD_{ \mu\nu}({\bf
k})\gamma_1^{\mu}\gamma_2^{\nu}+ V^V_{\rm conf}({\bf k})
\Gamma_1^{\mu}({\bf k})\Gamma_{2;\mu}(-{\bf k})+
 V^S_{\rm conf}({\bf k})\right],
\]

(b) for quark-diquark ($Qd$) interaction in the colour singlet state
\begin{eqnarray}
\label{dpot}
V({\bf p,q};M)&=&\frac{\langle d(P)|J_{\mu}|d(Q)\rangle}
{2\sqrt{E_d(p)E_d(q)}} \bar{u}_{Q}(p)  
\frac43\alpha_sD_{ \mu\nu}({\bf 
k})\gamma^{\nu}u_{Q}(q)\cr
&&+\psi^*_d(P)\bar u_Q(p)J_{d;\mu}\Gamma_Q^\mu({\bf k})
V_{\rm conf}^V({\bf k})u_{Q}(q)\psi_d(Q)\cr 
&&+\psi^*_d(P)
\bar{u}_{Q}(p)V^S_{\rm conf}({\bf k})u_{Q}(q)\psi_d(Q), 
\end{eqnarray}
where $\alpha_s$ is the QCD coupling constant, $\langle
d(P)|J_{\mu}|d(Q)\rangle$ is the vertex of the 
diquark-gluon interaction which takes into account the diquark
internal structure, $P=(E_d(p),-{\bf p})$, $Q=(E_d(q),-{\bf q})$ and $E_d(p)=\sqrt{{\bf p}^2+M_d^2}$. $D_{\mu\nu}$ is the  
gluon propagator in the Coulomb gauge, ${\bf k=p-q}$; $\gamma_{\mu}$ and $u(p)$ are 
the Dirac matrices and spinors
\begin{equation}
\label{spinor}
u^\lambda({p})=\sqrt{\frac{\epsilon(p)+m}{2\epsilon(p)}}
\left(
\begin{array}{c}1\cr {\displaystyle\frac{\bm{\sigma}
      {\bf  p}}{\epsilon(p)+m}}
\end{array}\right)\chi^\lambda,
\end{equation}
with $\epsilon(p)=\sqrt{{\bf p}^2+m^2}$.

The diquark state in the confining part of the quark-diquark
quasipotential (\ref{dpot}) is described by the wave functions
\begin{equation}
  \label{eq:ps}
  \psi_d(p)=\left\{\begin{array}{ll}1 &\qquad \text{ for the scalar diquark}\\
\varepsilon_d(p) &\qquad \text{ for the axial vector diquark}
\end{array}\right. ,
\end{equation}
where $\varepsilon_d$ is the polarization vector of the axial vector
diquark. The effective long-range vector vertex of the
diquark can be presented in the form  
\begin{equation}
  \label{eq:jc}
  J_{d;\mu}=\left\{\begin{array}{ll}
  \frac{\displaystyle (P+Q)_\mu}{\displaystyle
  2\sqrt{E_d(p)E_d(q)}}&\qquad \text{ for the scalar diquark}\cr
-\frac{\displaystyle (P+Q)_\mu}{\displaystyle2\sqrt{E_d(p)E_d(q)}}
  +\frac{\displaystyle i\mu_d}{\displaystyle 2M_d}\Sigma_\mu^\nu 
\tilde k_\nu
  &\qquad \text{ for the axial 
  vector diquark}\end{array}\right. ,
\end{equation}
where $\tilde k=(0,{\bf k})$. Here  $\Sigma_\mu^\nu$ is the antisymmetric tensor
\begin{equation}
  \label{eq:Sig}
  \left(\Sigma_{\rho\sigma}\right)_\mu^\nu=-i(g_{\mu\rho}\delta^\nu_\sigma
  -g_{\mu\sigma}\delta^\nu_\rho),
\end{equation}
and the axial vector diquark spin ${\bf S}_d$ is given by
$(S_{d;k})_{il}=-i\varepsilon_{kil}$. We choose the total
chromomagnetic moment of the axial vector 
diquark $\mu_d=0$ \cite{tetr}.

The effective long-range vector vertex of the quark is
defined by \cite{egf,sch}
\begin{equation}
\Gamma_{\mu}({\bf k})=\gamma_{\mu}+
\frac{i\kappa}{2m}\sigma_{\mu\nu}\tilde k^{\nu}, \qquad \tilde
k=(0,{\bf k}),
\end{equation}
where $\kappa$ is the Pauli interaction constant characterizing the
anomalous chromomagnetic moment of quarks. In the configuration space
the vector and scalar confining potentials in the nonrelativistic
limit reduce to
\begin{eqnarray}
V^V_{\rm conf}(r)&=&(1-\varepsilon)V_{\rm conf}(r),\nonumber\\
V^S_{\rm conf}(r)& =&\varepsilon V_{\rm conf}(r),
\end{eqnarray}
with 
\begin{equation}
V_{\rm conf}(r)=V^S_{\rm conf}(r)+
V^V_{\rm conf}(r)=Ar+B,
\end{equation}
where $\varepsilon$ is the mixing coefficient.

The constituent quark masses $m_u=m_d=0.33$ GeV, $m_s=0.5$ GeV,
$m_c=1.55$ GeV, $m_b=4.88$ GeV,  and 
the parameters of the linear potential $A=0.18$ GeV$^2$ and $B=-0.3$ GeV
have the usual values of quark models.  The value of the mixing
coefficient of vector and scalar confining potentials $\varepsilon=-1$
has been determined from the consideration of charmonium radiative
decays \cite{efg} and the heavy quark expansion \cite{fg}. 
Finally, the universal Pauli interaction constant $\kappa=-1$ has been
fixed from the analysis of the fine splitting of heavy quarkonia ${
}^3P_J$- states \cite{efg}.  Note that the 
long-range chromomagnetic contribution to the potential in our model
is proportional to $(1+\kappa)$ and thus vanishes for the 
chosen value of $\kappa=-1$.

Since we deal with diquarks and baryons containing light quarks we adopt for the QCD
coupling constant $\alpha_s(\mu^2)$ the
simplest model with freezing \cite{bvb}, namely
\begin{equation}
  \label{eq:alpha}
  \alpha_s(\mu^2)=\frac{4\pi}{\displaystyle\beta_0
\ln\frac{\mu^2+M_B^2}{\Lambda^2}}, \qquad \beta_0=11-\frac23n_f,
\end{equation}
where the scale is taken as $\mu=2m_1
m_2/(m_1+m_2)$, the background mass is $M_B=2.24\sqrt{A}=0.95$~GeV \cite{bvb}, and
$\Lambda=413$~MeV was fixed from fitting the $\rho$
mass \cite{lregge}. Note that an other popular
parametrization of $\alpha_s$ with freezing \cite{shirkov} leads to close
values.  

\subsection{Light diquarks}
\label{sec:ld}

At the first step, we present the masses and form factors of the light
diquark \cite{hbar}. As it is well known, the light quarks are highly
relativistic, which makes the $v/c$ expansion inapplicable and thus,
a completely relativistic treatment is required. To achieve this goal in
describing light 
diquarks, we closely follow our consideration of the light
meson spectra  \cite{lmes} and adopt the same procedure to make the relativistic
quark potential local by replacing
$\epsilon_{1,2}(p)=\sqrt{m_{1,2}^2+{\bf p}^2}\to E_{1,2}$  
(see (\ref{ee}) and discussion in Ref.~\cite{lmes}). 

The quasipotential equation (\ref{quas}) is solved numerically for the
complete relativistic potential  which depends on the
diquark mass in a complicated highly nonlinear way \cite{hbar}.  The obtained
ground state masses of scalar  and axial
vector  light diquarks are presented in
Table~\ref{tab:dmass}.  
 
In order to determine the diquark interaction with the gluon field  $\langle
d(P)|J_{\mu}|d(Q)\rangle$, which
takes into account the diquark structure, it is
necessary to calculate the corresponding matrix element of the quark
current between diquark states.
This diagonal matrix element can be
parametrized by the set of elastic form factors in the  following way

(a) scalar diquark ($d=S$)
\begin{equation}
  \label{eq:sff}
  \langle S(P)\vert J_\mu \vert S(Q)\rangle=h_+(k^2)(P+Q)_\mu,
\end{equation}

(b) axial vector diquark ($d=A$) 
\begin{eqnarray}
  \label{eq:avff}
\langle A(P)\vert J_\mu \vert A(Q)\rangle&=&
-[\varepsilon_d^*(P)\cdot\varepsilon_d(Q)]h_1(k^2)(P+Q)_\mu\cr
&&+h_2(k^2)
\left\{[\varepsilon_d^*(P) \cdot Q]\varepsilon_{d;\mu}(Q)+
  [\varepsilon_d(Q) \cdot P] 
\varepsilon^*_{d;\mu}(P)\right\}\cr
&&+h_3(k^2)\frac1{M_{A}^2}[\varepsilon^*_d(P) \cdot Q]
    [\varepsilon_d(Q) \cdot P](P+Q)_\mu, 
\end{eqnarray}
where $k=P-Q$ and $\varepsilon_d(P)$ is the polarization vector of the
axial vector diquark. 

The calculation of the
matrix element of the quark current $J_\mu=\bar q
\gamma^\mu q$  between the diquark states leads to the
emergence of the form factor $F(r)$ entering the vertex of the
diquark-gluon interaction \cite{hbar}. 
Then the elastic form factors in Eqs.~(\ref{eq:sff}) and (\ref{eq:avff}) are
expressed by
\begin{eqnarray*}
  h_+(k^2)&=&h_1(k^2)=h_2(k^2)=F({\bf k}^2),\cr
h_3(k^2)&=&0,
\end{eqnarray*}
where the form factor $F(r)$ is given by
the overlap integral of the diquark wave functions.
Using the numerical diquark wave functions we find that  $F(r)$ can be
approximated  with high accuracy by the expression \cite{hbar} 
\begin{equation}
  \label{eq:fr}
  F(r)=1-e^{-\xi r -\zeta r^2}.
\end{equation}
The values of the parameters $\xi$ and $\zeta$ for the  ground states
of the  scalar $[q,q']$ and axial vector $\{q,q'\}$
light diquarks are given in Table~\ref{tab:dmass}. 
\begin{table}
  \caption{Masses $M$ and form factor  parameters of
    diquarks. $S$ and $A$ 
    denote scalar and axial vector diquarks which are antisymmetric $[\cdots]$ and
    symmetric $\{\cdots\}$ in flavour, respectively \cite{hbar}. }
  \label{tab:dmass}
\begin{ruledtabular}
\begin{tabular}{ccccc}
Quark& Diquark&   $M$ &$\xi$ & $\zeta$
 \\
content &type & (MeV)& (GeV)& (GeV$^2$)  \\
\hline
$[u,d]$&S & 710 & 1.09 & 0.185  \\
$\{u,d\}$&A & 909 &1.185 & 0.365  \\
$[u,s]$& S & 948 & 1.23 & 0.225 \\
$\{u,s\}$& A & 1069 & 1.15 & 0.325\\
$\{s,s\}$ & A& 1203 & 1.13 & 0.280\\
  \end{tabular}
\end{ruledtabular}
\end{table}

\subsection{Heavy baryons as heavy-quark--light-diquark bound systems}

At the second step, we calculate the masses of heavy baryons as the bound
states of a heavy quark and light diquark. Since we are
considering highly excited heavy baryons,  we do not expand the
potential of the heavy-quark--light-diquark interaction (\ref{dpot})
either in $p/m_Q$ or in $p/m_d$ and treat both light diquark and heavy
quark fully relativistically.  To simplify the potential and to make
it local in configuration space we follow the
same procedure, which was used for light quarks in a diquark, and  replace
in  Eqs.~(\ref{dpot}), (\ref{spinor}), (\ref{eq:jc}):

(a) the diquark energies 
$$E_d(p)\equiv\sqrt{{\bf p}^2+M_d^2}\to
E_d=\frac{M^2-m_Q^2+M_d^2}{2M},$$ 

(b) the heavy quark energies 
$$\epsilon_Q(p)\equiv\sqrt{{\bf p}^2+m_Q^2}\to
E_Q=\frac{M^2-M_d^2+m_Q^2}{2M}.$$ 
These substitutions make the Fourier transform of the potential
(\ref{dpot}) local, but introduce a complicated nonlinear dependence of
the potential on the baryon mass $M$ through the on-mass-shell energies
$E_d$ and $E_Q$. 

The resulting $Q\bar d$ potential then reads
\begin{equation}
  \label{eq:v}
  V(r)= V_{\rm SI}(r)+ V_{\rm SD}(r),
\end{equation}
where  the spin-independent $V_{\rm SI}(r)$ part is given by
\begin{eqnarray}
\label{si}\!\!\!\!\!\!\!
V_{\rm SI}(r)&=&\hat V_{\rm Coul}(r)
+V_{\rm  conf}(r)+\frac{1}{E_dE_{Q}}\Bigg\{ \frac12(E_Q^2-m_Q^2+E_d^2-M_d^2)\left[\hat
  V_{\rm Coul}(r)+V^V_{\rm conf}(r)\right]\cr
&&+\frac{1}{4}\Delta \left[2V_{\rm Coul}(r)+V^V_{\rm conf}(r)\right]+\hat V'_{\rm Coul}(r)\frac{{\bf L}^2}{2r}\Bigg\}+\frac1{E_Q(E_Q+m_Q)}\Biggl\{
-(E_Q^2-m_Q^2)V^S_{\rm conf}(r)\cr
&&+\frac14\Delta\left(\hat V_{\rm
Coul}(r)-V_{\rm conf}(r)-2\left[\frac{E_Q-m_Q}{2m_Q}-(1+\kappa)\frac{E_Q+m_Q}{2m_Q}\right]V^V_{\rm conf}(r)
\right)\Biggr\}.
\end{eqnarray}
 Here $\Delta$ is the Laplace operator, and $\hat V_{\rm Coul}(r)$ is the smeared Coulomb
potential which accounts for the diquark internal structure
$$\hat V_{\rm Coul}(r)=-\frac43\alpha_s\frac{ F(r)}{r}.$$

The structure of the spin-dependent potential is given by
\begin{equation}
  \label{eq:vsd}
   V_{\rm SD}(r)=a_1\, {\bf L}{\bf S}_d+a_2\, {\bf L}{\bf S}_Q+
b \left[-{\bf S}_d{\bf S}_Q+\frac3{r^2}({\bf S}_d{\bf r})({\bf
    S}_Q{\bf r})\right]+ c\, {\bf S}_d{\bf S}_Q,
\end{equation}
where $ {\bf L}$ is the orbital angular momentum; ${\bf S}_d$ and
${\bf S}_Q$ are the diquark and quark spin operators, respectively. 
The coefficients $a_1$, $a_2$, $b$ and $c$ are expressed
through the corresponding derivatives of the smeared Coulomb and
confining potentials:
\begin{eqnarray}
\label{a1}
 a_1&= &\frac{1}{M_d(E_d+M_d)}\frac1{r}\Biggl[\frac{M_d}{E_d}
\hat  V'_{\rm Coul}(r)-V'_{\rm conf}(r)+\mu_d\frac{E_d+M_d}{2M_d}V'^V_{\rm conf}(r)\Biggr]\cr 
&& +\frac1{E_dE_Q}\frac1{r}\Biggl[\left(\hat V'_{\rm Coul}(r)+\frac{\mu_d}2\frac{E_d}{M_d}V'^V_{\rm
    conf}(r)\right)+\frac{E_d}{M_d}\left(\frac{E_d-M_d}{E_Q+m_Q}
+\frac{E_Q-m_Q}{E_d+M_d}\right) V'^S_{\rm conf}(r) 
\Biggr],\ \ \ \ \
\end{eqnarray}
\begin{eqnarray}
\label{a2}
a_2&=&\frac{1}{E_dE_{Q}}\frac1{r}\Biggl\{\hat V'_{\rm Coul}(r)
-\left[\frac{E_Q-m_Q}{2m_Q}- 
 (1+\kappa)\frac{E_Q+m_Q}{2m_Q}\right]V'^V_{\rm conf}(r)\Biggr\} \cr
&&+\frac{1}{E_Q(E_Q+m_Q)}\frac1{r}\Biggl\{\hat V'_{\rm Coul}(r)-
V'_{\rm conf}(r)
-2\left[\frac{E_Q-m_Q}{2m_Q}- 
 (1+\kappa)\frac{E_Q+m_Q}{2m_Q}\right]V'^V_{\rm conf}(r)\Biggr\},\cr
&&
\end{eqnarray}
\begin{eqnarray}
\label{b}
b&=&\frac13\frac{1}{E_dE_{Q}}\Biggl\{\frac1{r}\hat V'_{\rm Coul}(r)-\hat V''_{\rm
    Coul}(r)\cr
&&-\frac{\mu_d}2\frac{E_d}{M_d}\left[\frac{E_Q-m_Q}{2m_Q}- 
 (1+\kappa)\frac{E_Q+m_Q}{2m_Q}\right]\left[\frac1{r}V'^V_{\rm
      conf}(r)- V''^V_{\rm conf}(r)\right]\Biggr\},
\end{eqnarray}
\begin{eqnarray}
\label{c}
c&=&\frac23\frac{1}{E_dE_{Q}}\left\{\Delta \hat V_{\rm Coul}(r)-
\frac{\mu_d}2\frac{E_d}{M_d}\left[\frac{E_Q-m_Q}{2m_Q}- 
 (1+\kappa)\frac{E_Q+m_Q}{2m_Q}\right]\Delta V^V_{\rm 
conf}(r)\right\}.
\end{eqnarray}
Both the one-gluon exchange and confining potential contribute
to the quark-diquark spin-orbit interaction. The quasipotential
(\ref{eq:v})-(\ref{c}) generalizes the one obtained previously
in the framework of the heavy quark $1/m_Q$ expansion
\cite{excbar}. Note that the expansion of the extended
potential (\ref{eq:v})--(\ref{c}) up to the second order in $1/m_Q$
and the subsequent substitution of the quark energies $\epsilon_Q(p)$
by the corresponding energies on mass shell $E_Q$, reproduces  the potential
of Ref.~\cite{excbar}.   

For the scalar diquark (${S}_d=0$) only the term (\ref{a2}), responsible for the heavy quark
spin-orbit interaction, contributes to the spin-dependent potential
(\ref{eq:vsd}), whereas for the axial-vector diquark (${S}_d=1$) all
terms (\ref{a1})--(\ref{c}) contribute to the
spin-dependent potential (\ref{eq:vsd}).
 Solving numerically  Eq.~(\ref{quas}) with the complete relativistic
 quasipotential (\ref {eq:v}) we  get 
the baryon wave function $\Psi_B$. Then the total baryon wave function 
is a product of $\Psi_B$ and the spin function $U_B$ (for
details see Eq. (43) of Ref.~\cite{dhbd}).

It is necessary to note that the presence of the spin-orbit
interaction ${\bf L}{\bf S}_Q$ and of the tensor interaction
in the quark-diquark potential 
(\ref{a1})--(\ref{b}) results in a mixing of states which
have the same total angular momentum $J$ and parity $P$ but different light
diquark total angular momentum  (${\bf L}+{\bf S}_d$). Such mixing is
considered along the same lines 
as in our previous calculations of the mass spectra of doubly heavy
baryons \cite{efgm}.

\section{Results and discussion}
\label{sec:rd}

\subsection{Heavy baryon masses}
\label{sec:hbm}

We solve numerically the quasipotential equation with the
quasipotential (\ref{eq:v}) which nonperturbatively accounts for the
relativistic dynamics both of the light diquark $d$ and heavy quark $Q$.   
The calculated values of the ground and excited state baryon masses are given in
Tables~\ref{tab:lq}-\ref{tab:om} in comparison with available
experimental data \cite{pdg}.
In the first two columns we give the baryon quantum numbers ($I(J^P)$) and the 
state of the heavy-quark--light-diquark ($Qd$) bound system (in usual
notations $(n_r+1)L$), while in the remaining
columns our predictions for the masses and experimental data are
shown. 

It is important to note that in the adopted quark-diquark picture of
heavy baryons we consider solely the orbital and radial excitations
between the heavy quark and light diquark, while light diquarks are
taken in the ground (scalar or axial-vector) state. As a result,
we get significantly less excited states than in the genuine
three-quark picture of a baryon. As it is seen from
Tables~\ref{tab:lq}-\ref{tab:om}, such an approach  is supported by available
experimental data, which are nicely accommodated in the quark-diquark
picture.     

Comparing the new values of heavy baryon masses presented in
Tables~\ref{tab:lq}-\ref{tab:om} with the previous results, obtained by using
the heavy quark expansion \cite{excbar}, we can estimate the importance
of higher order corrections in $1/m_Q$. Such comparison confirms
expectations that they are mainly important for highly
excited heavy baryon states and that charmed baryons are stronger affected
than the bottom ones. Indeed, the difference of masses, obtained with
and without expansion in $1/m_Q$, does not
exceed a few MeV
for the ground state heavy baryons, while for excited states such
difference in some cases reaches tens MeV, especially for the charmed
baryons. 

%\newpage
\begin{table}
\caption{\label{tab:lq} Masses of the $\Lambda_Q$ ($Q=c,b$) heavy baryons (in MeV).}
\begin{ruledtabular}
\begin{tabular}{cccccc}
& & \multicolumn{2}{c}{\hspace{-0.8cm}\underline{\hspace{2cm}$Q=c$\hspace{2cm}}}\hspace{-0.8cm}& \multicolumn{2}{c}{\hspace{-1cm}\underline{\hspace{1.5cm}$Q=b$\hspace{1.8cm}}}\hspace{-0.8cm}\\
$I(J^P)$& $Qd$ state & $M$ & $M^{\rm exp}$ \cite{pdg}&  $M$ & $M^{\rm
  exp}$  \cite{pdg}\\
\hline
$0(\frac12^+)$& $1S$ & 2286 & 2286.46(14) & 5620 & 5620.2(1.6)\\
$0(\frac12^+)$& $2S$ & 2769 & 2766.6(2.4)?& 6089 \\
$0(\frac12^+)$& $3S$ & 3130 &           & 6455  \\
$0(\frac12^+)$& $4S$ & 3437 &           & 6756  \\
$0(\frac12^+)$& $5S$ & 3715 &           & 7015  \\
$0(\frac12^+)$& $6S$ & 3973 &           & 7256  \\
$0(\frac12^-)$& $1P$ & 2598 & 2595.4(6) & 5930  \\
$0(\frac12^-)$& $2P$ & 2983 & 2939.3$(^{1.4}_{1.5})$? & 6326  \\
$0(\frac12^-)$& $3P$ & 3303 &           & 6645  \\
$0(\frac12^-)$& $4P$ & 3588 &           & 6917  \\
$0(\frac12^-)$& $5P$ & 3852 &           & 7157  \\
$0(\frac32^-)$& $1P$ & 2627 & 2628.1(6) & 5942  \\
$0(\frac32^-)$& $2P$ & 3005 &           & 6333  \\
$0(\frac32^-)$& $3P$ & 3322 &           & 6651  \\
$0(\frac32^-)$& $4P$ & 3606 &           & 6922  \\
$0(\frac32^-)$& $5P$ & 3869 &           & 7171  \\
$0(\frac32^+)$& $1D$ & 2874 &           & 6190  \\
$0(\frac32^+)$& $2D$ & 3189 &           & 6526  \\
$0(\frac32^+)$& $3D$ & 3480 &           & 6811  \\
$0(\frac32^+)$& $4D$ & 3747 &           & 7060  \\
$0(\frac52^+)$& $1D$ & 2880 & 2881.53(35)& 6196  \\
$0(\frac52^+)$& $2D$ & 3209 &           & 6531 \\
$0(\frac52^+)$& $3D$ & 3500 &           & 6814 \\
$0(\frac52^+)$& $4D$ & 3767 &           & 7063 \\
$0(\frac52^-)$& $1F$ & 3097 &           & 6408  \\
$0(\frac52^-)$& $2F$ & 3375 &           & 6705  \\
$0(\frac52^-)$& $3F$ & 3646 &           & 6964  \\
$0(\frac52^-)$& $4F$ & 3900 &           & 7196  \\
$0(\frac72^-)$& $1F$ & 3078 &           & 6411  \\
$0(\frac72^-)$& $2F$ & 3393 &           & 6708  \\
$0(\frac72^-)$& $3F$ & 3667&           & 6966  \\
$0(\frac72^-)$& $4F$ & 3922 &           & 7197  \\
$0(\frac72^+)$& $1G$ & 3270 &           & 6598  \\
$0(\frac72^+)$& $2G$ & 3546 &           & 6867  \\
$0(\frac92^+)$& $1G$ & 3284 &           & 6599  \\
$0(\frac92^+)$& $2G$ & 3564 &           & 6868  \\
$0(\frac92^-)$& $1H$ & 3444 &           & 6767  \\
$0(\frac{11}2^-)$& $1H$ & 3460 &           & 6766  \\
\end{tabular}
\end{ruledtabular}
\end{table}
\newpage
\begin{longtable}[h!]{@{\ \ \ }c@{ \ \ \ \ \ \ \ }c@{ \
      \ \ \ \ \  \ \ }c@{\ \ \ \ \ \ \ \
        \ \ \ \ \ }c@{\ \ \ \ \ \ \ \ \ \
      \ \  \ \  }c@{\ \ \ \ \ \ \ \ \ \ \ \ \ }c@{\ \ \  }}%{ccccccccc}
\caption{Masses of the $\Sigma_Q$ ($Q=c,b$) 
heavy baryons (in MeV).}
\label{tab:sq}\vspace*{-0.5cm}\\
\hline
\hline
\\[1pt]
%\begin{ruledtabular}
%\begin{tabular}{ccccccccc}
& & \multicolumn{2}{c}{\hspace{-1.9cm}\underline{\hspace{2.2cm}$Q=c$\hspace{2.2cm}}}\hspace{-1cm}& \multicolumn{2}{c}{\hspace{-1.2cm}\underline{\hspace{1.9cm}$Q=b$\hspace{1.9cm}}}\hspace{-1cm}\\
$I(J^P)$& $Qd$ state & $M$ & $M^{\rm exp}$  \cite{pdg}  &  $M$ &
$M^{\rm exp}$ \cite{pdg}\\[2pt]
\hline
\endfirsthead
\caption[]{(continued)}\vspace*{-0.5cm}\\
\hline\hline\\[1pt]
& & \multicolumn{2}{c}{\hspace{-1.9cm}\underline{\hspace{2.2cm}$Q=c$\hspace{2.2cm}}}\hspace{-1cm}& \multicolumn{2}{c}{\hspace{-1.2cm}\underline{\hspace{1.9cm}$Q=b$\hspace{1.9cm}}}\hspace{-1cm}\\
$I(J^P)$& $Qd$ state & $M$ & $M^{\rm exp}$  \cite{pdg}  &  $M$ &
$M^{\rm exp}$ \cite{pdg}\\[2pt]
\hline
\endhead
\\[2pt]\hline
\hline
\endfoot
\endlastfoot
$1(\frac12^+)$& $1S$ & 2443 & 2453.76(18) & 5808 &
5807.8(2.7)\\
$1(\frac12^+)$& $2S$ & 2901 &            & 6213&  \\
$1(\frac12^+)$& $3S$ & 3271 &            & 6575& \\
$1(\frac12^+)$& $4S$ & 3581 &            & 6869&  \\
$1(\frac12^+)$& $5S$ & 3861 &            & 7124&  \\
$1(\frac32^+)$& $1S$ & 2519 & 2518.0(5)  & 5834 &
5829.0(3.4)\\
$1(\frac32^+)$& $2S$ & 2936 &   2939.3$(^{1.4}_{1.5})$? &
6226 &\\
$1(\frac32^+)$& $3S$ & 3293 &            &  6583 &\\
$1(\frac32^+)$& $4S$ & 3598 &            &  6876 &\\
$1(\frac32^+)$& $5S$ & 3873 &            &  7129 &\\
$1(\frac12^-)$& $1P$ & 2799 & 2802($^4_7$)& 6101 &     \\
$1(\frac12^-)$& $2P$ & 3172 &            &  6440 & \\
$1(\frac12^-)$& $3P$ & 3488 &            &  6756 & \\
$1(\frac12^-)$& $4P$ & 3770 &            &  7024 & \\
$1(\frac12^-)$& $1P$ & 2713 &            &  6095 &    \\
$1(\frac12^-)$& $2P$ & 3125 &            &  6430 & \\
$1(\frac12^-)$& $3P$ & 3455 &            &  6742 & \\
$1(\frac12^-)$& $4P$ & 3743 &            &  7008 & \\
$1(\frac32^-)$& $1P$ & 2798 & 2802($^4_7$)&  6096 & \\
$1(\frac32^-)$& $2P$ & 3172 &            &  6430 & \\
$1(\frac32^-)$& $3P$ & 3486 &            &  6742 & \\
$1(\frac32^-)$& $4P$ & 3768 &            &  7009 & \\
$1(\frac32^-)$& $1P$ & 2773 & 2766.6(2.4)?& 6087 & \\
$1(\frac32^-)$& $2P$ & 3151 &            &  6423 & \\
$1(\frac32^-)$& $3P$ & 3469 &            &  6736 & \\
$1(\frac32^-)$& $4P$ & 3753 &            &  7003 & \\
$1(\frac52^-)$& $1P$ & 2789 &            &  6084 & \\
$1(\frac52^-)$& $2P$ & 3161 &            &  6421 & \\
$1(\frac52^-)$& $3P$ & 3475 &            &  6732 & \\
$1(\frac52^-)$& $4P$ & 3757 &            &  6999 & \\
$1(\frac12^+)$& $1D$ & 3041 &            &  6311 & \\
$1(\frac12^+)$& $2D$ & 3370 &            &  6636 & \\
$1(\frac32^+)$& $1D$ & 3043 &            &  6326 & \\
$1(\frac32^+)$& $2D$ & 3366 &            &  6647 & \\
$1(\frac32^+)$& $1D$ & 3040 &            &  6285 & \\
$1(\frac32^+)$& $2D$ & 3364 &            &  6612 & \\
$1(\frac52^+)$& $1D$ & 3038 &            &  6284 & \\
$1(\frac52^+)$& $2D$ & 3365 &            &  6612 & \\
$1(\frac52^+)$& $1D$ & 3023 &            &  6270 & \\
$1(\frac52^+)$& $2D$ & 3349 &            &  6598 & \\
$1(\frac72^+)$& $1D$ & 3013 &            &  6260 & \\
$1(\frac72^+)$& $2D$ & 3342 &            &  6590 & \\%\pagebreak
$1(\frac32^-)$& $1F$ & 3288 &            &  6550 & \\
$1(\frac52^-)$& $1F$ & 3283 &            &  6564 & \\
$1(\frac52^-)$& $1F$ & 3254 &            &  6501 & \\
$1(\frac72^-)$& $1F$ & 3253 &            &  6500 & \\
$1(\frac72^-)$& $1F$ & 3227 &            &  6472 & \\
$1(\frac92^-)$& $1F$ & 3209 &            &  6459 & \\
$1(\frac52^+)$& $1G$ & 3495 &            &  6749 & \\
$1(\frac72^+)$& $1G$ & 3483 &            &  6761 & \\
$1(\frac72^+)$& $1G$ & 3444 &            &  6688 & \\
$1(\frac92^+)$& $1G$ & 3442 &            &  6687 & \\
$1(\frac92^+)$& $1G$ & 3410 &            &  6648 & \\
$1(\frac{11}2^+)$& $1G$ &3386 &          &  6635 & \\%[2pt]
\hline\hline
\end{longtable}
%\end{ruledtabular}
% \end{table}

\begin{table}%[h]
\caption{\label{tab:xs} Masses of the $\Xi_Q$ ($Q=c,b$) heavy baryons
  with the scalar diquark (in MeV).}
\begin{ruledtabular}
\begin{tabular}{cccccc}
& & \multicolumn{2}{c}{\hspace{-0.8cm}\underline{\hspace{2cm}$Q=c$\hspace{2cm}}}\hspace{-0.8cm}& \multicolumn{2}{c}{\hspace{-1cm}\underline{\hspace{1.5cm}$Q=b$\hspace{1.8cm}}}\hspace{-0.8cm}\\
% & & \multicolumn{2}{c}{\hspace{-0.5cm}\underline{\hspace{2.2cm}$Q=c$\hspace{2.2cm}}}\hspace{0.cm}& \multicolumn{2}{c}{\hspace{-0.5cm}\underline{\hspace{2.2cm}$Q=b$\hspace{2.2cm}}}\hspace{-0.2cm}\\
$I(J^P)$& $Qd$ state & $M$ & $M^{\rm exp}$ \cite{pdg} & $M$& $M^{\rm exp}$ \cite{pdg}  \\
\hline
$\frac12(\frac12^+)$& $1S$ & 2476 & 2470.88$(^{34}_{80})$   & 5803& 5790.5(2.7) \\
$\frac12(\frac12^+)$& $2S$ & 2959 &             & 6266  \\
$\frac12(\frac12^+)$& $3S$ & 3323 &             & 6601  \\
$\frac12(\frac12^+)$& $4S$ & 3632 &             & 6913  \\
$\frac12(\frac12^+)$& $5S$ & 3909 &             & 7165  \\
$\frac12(\frac12^+)$& $6S$ & 4166 &             & 7415  \\
$\frac12(\frac12^-)$& $1P$ & 2792 & 2791.8(3.3) & 6120  \\
$\frac12(\frac12^-)$& $2P$ & 3179 &             & 6496  \\
$\frac12(\frac12^-)$& $3P$ & 3500 &             & 6805  \\
$\frac12(\frac12^-)$& $4P$ & 3785 &             & 7068  \\
$\frac12(\frac12^-)$& $5P$ & 4048 &             & 7302  \\
$\frac12(\frac32^-)$& $1P$ & 2819 & 2819.6(1.2) & 6130  \\
$\frac12(\frac32^-)$& $2P$ & 3201 &             & 6502  \\
$\frac12(\frac32^-)$& $3P$ & 3519 &             & 6810  \\
$\frac12(\frac32^-)$& $4P$ & 3804 &             & 7073  \\
$\frac12(\frac32^-)$& $5P$ & 4066 &             & 7306  \\
$\frac12(\frac32^+)$& $1D$ & 3059 &3054.2(1.3)  & 6366  \\
$\frac12(\frac32^+)$& $2D$ & 3388 &             & 6690  \\
$\frac12(\frac32^+)$& $3D$ & 3678 &             & 6966  \\
$\frac12(\frac32^+)$& $4D$ & 3945 &             & 7208  \\
$\frac12(\frac52^+)$& $1D$ & 3076 & 3079.9(1.4) & 6373  \\
$\frac12(\frac52^+)$& $2D$ & 3407 &             & 6696 \\
$\frac12(\frac52^+)$& $3D$ & 3699 &             & 6970 \\
$\frac12(\frac52^+)$& $4D$ & 3965 &             & 7212 \\
$\frac12(\frac52^-)$& $1F$ & 3278 &             & 6577  \\
$\frac12(\frac52^-)$& $2F$ & 3575 &             & 6863  \\
$\frac12(\frac52^-)$& $3F$ & 3845 &             & 7114  \\
$\frac12(\frac52^-)$& $4F$ & 4098 &             & 7339  \\
$\frac12(\frac72^-)$& $1F$ & 3292 &             & 6581  \\
$\frac12(\frac72^-)$& $2F$ & 3592 &             & 6867  \\
$\frac12(\frac72^-)$& $3F$ & 3865 &             & 7117  \\
$\frac12(\frac72^-)$& $4F$ & 4120 &             & 7342  \\
$\frac12(\frac72^+)$& $1G$ & 3469 &           & 6760  \\
$\frac12(\frac72^+)$& $2G$ & 3745 &           & 7020  \\
$\frac12(\frac92^+)$& $1G$ & 3483 &           & 6762  \\
$\frac12(\frac92^+)$& $2G$ & 3763 &           & 7032  \\
$\frac12(\frac92^-)$& $1H$ & 3643 &           & 6933  \\
$\frac12(\frac{11}2^-)$& $1H$ & 3658 &           & 6934  \\
\end{tabular}
\end{ruledtabular}
\end{table}

\subsection{Regge trajectories of heavy baryons}
\label{sec:rthb}

In the presented analysis we calculated masses of both orbitally and radially excited
heavy baryons up to rather high excitation numbers ($L=5$ and
$n_r=5$). This makes it possible 
to construct the heavy baryon Regge trajectories both in the
$(J,M^2)$ and in the $(n_r,M^2)$ planes. We use the following  definitions. \\ 
(a) The $(J,M^2)$ Regge trajectory:

\begin{equation}
  \label{eq:reggej}
J=\alpha M^2+\alpha_0;
\end{equation}

\noindent (b) The $(n_r,M^2)$ Regge trajectory:

\begin{equation}
  \label{eq:reggen}
n_r=\beta M^2+\beta_0,
\end{equation}
where $\alpha$, $\beta$ are the slopes and  $\alpha_0$, $\beta_0$ are
intercepts.

\begin{figure}[htb]
 \includegraphics[width=8cm]{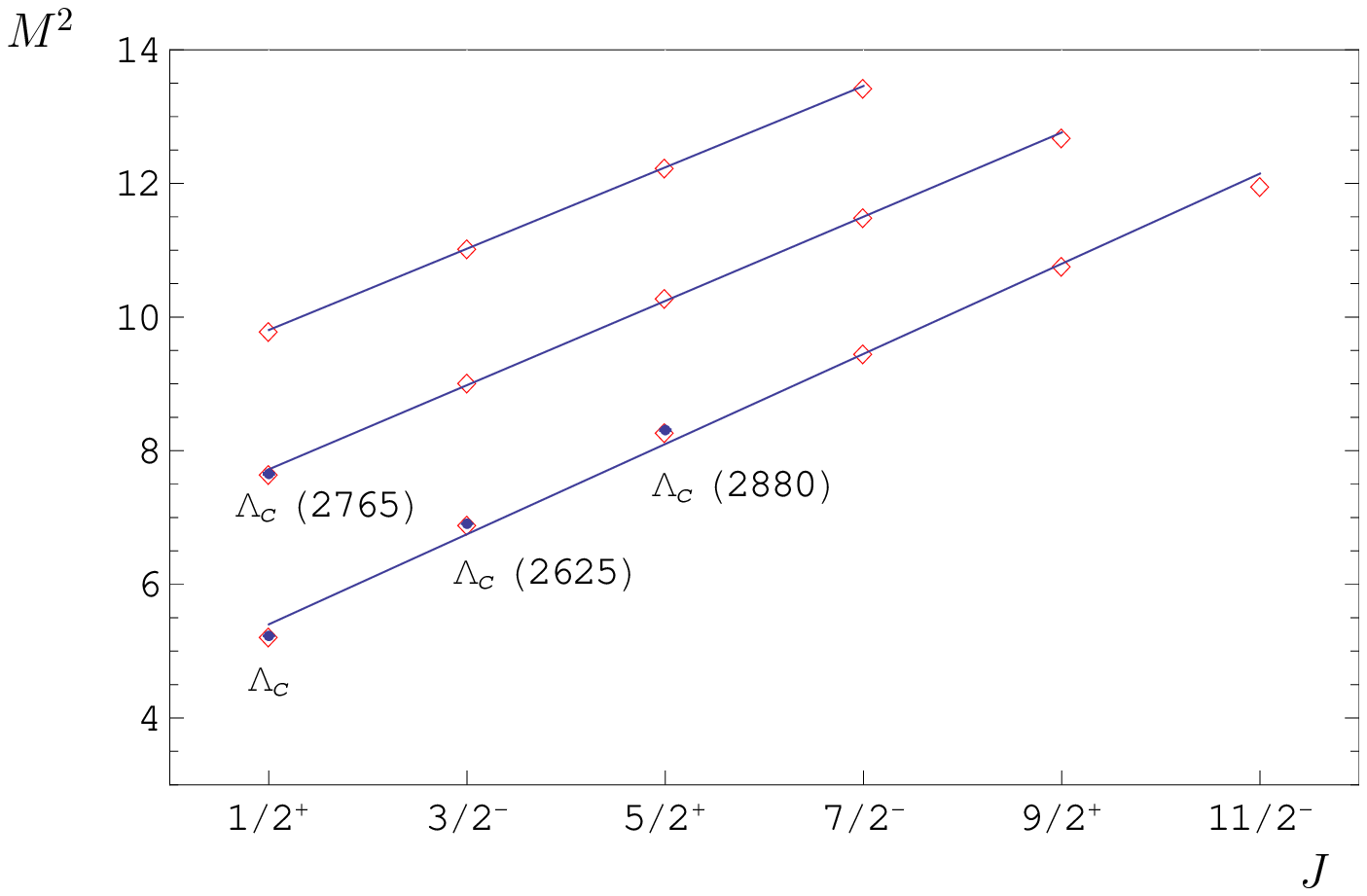}\ \ \ \includegraphics[width=8cm]{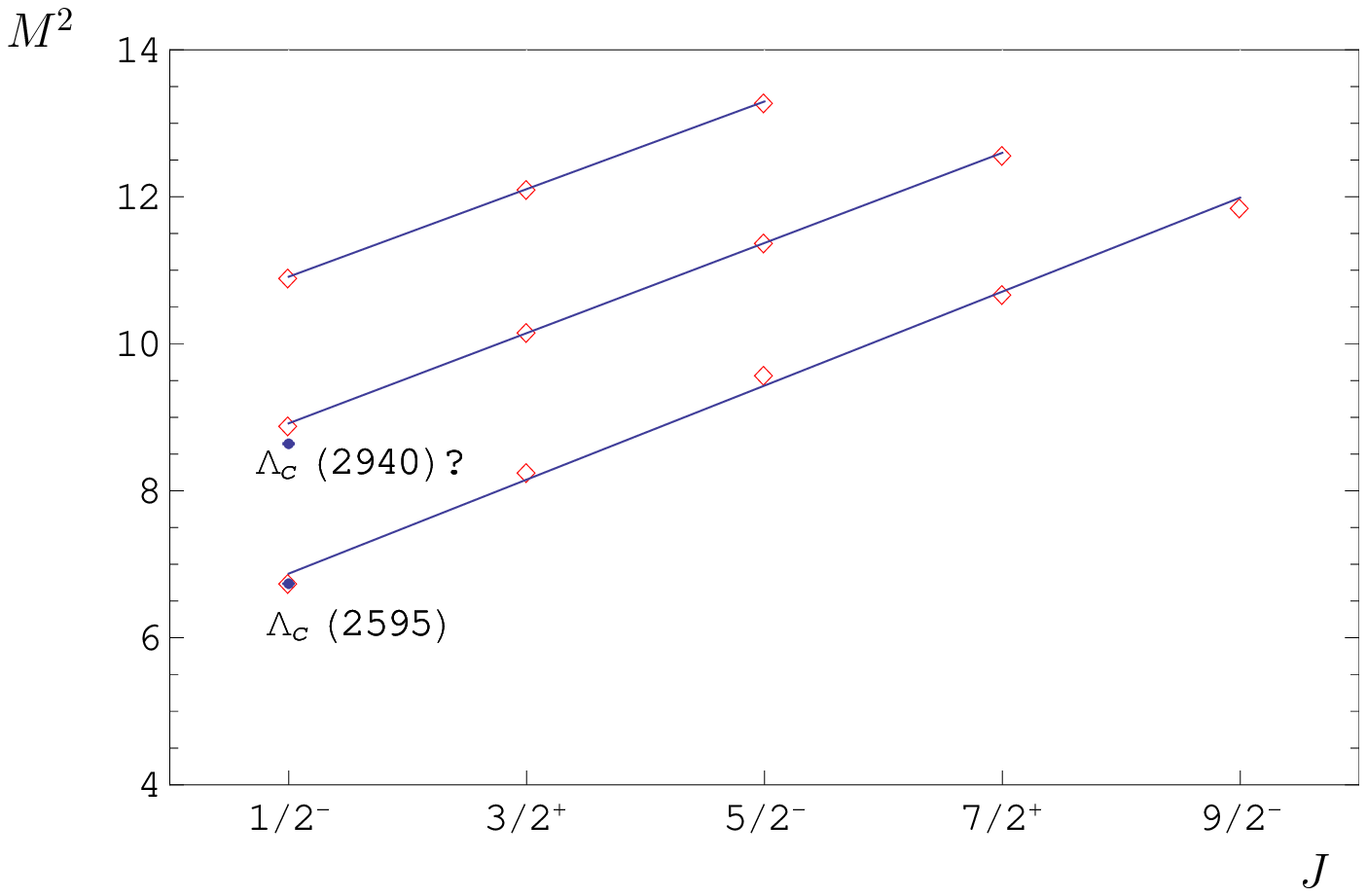}\\
\vspace*{-0.3cm}\hspace*{0.5cm} (a) \hspace*{7.5cm} (b) 
\caption{\label{fig:lambda_j} Parent and daughter ($J, M^2$) Regge trajectories for
  the $\Lambda_c$ baryons with natural (a) and unnatural (b) parities. Diamonds are predicted
  masses. Available experimental data are given by dots with  particle
  names; $M^2$ is in GeV$^2$. } 
\end{figure}

\begin{figure}
 \includegraphics[width=8.cm]{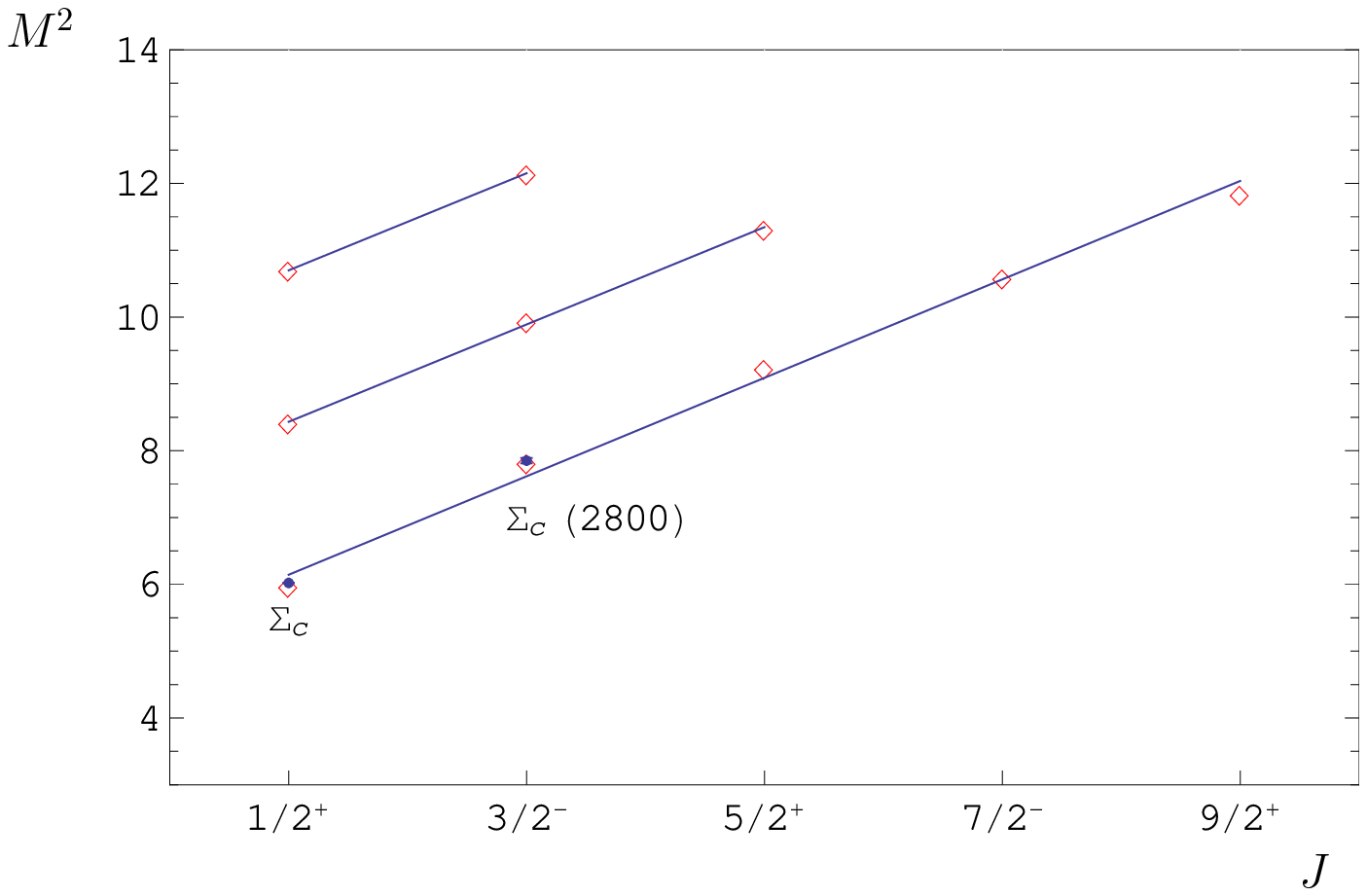}\ \
 \ \includegraphics[width=8.cm]{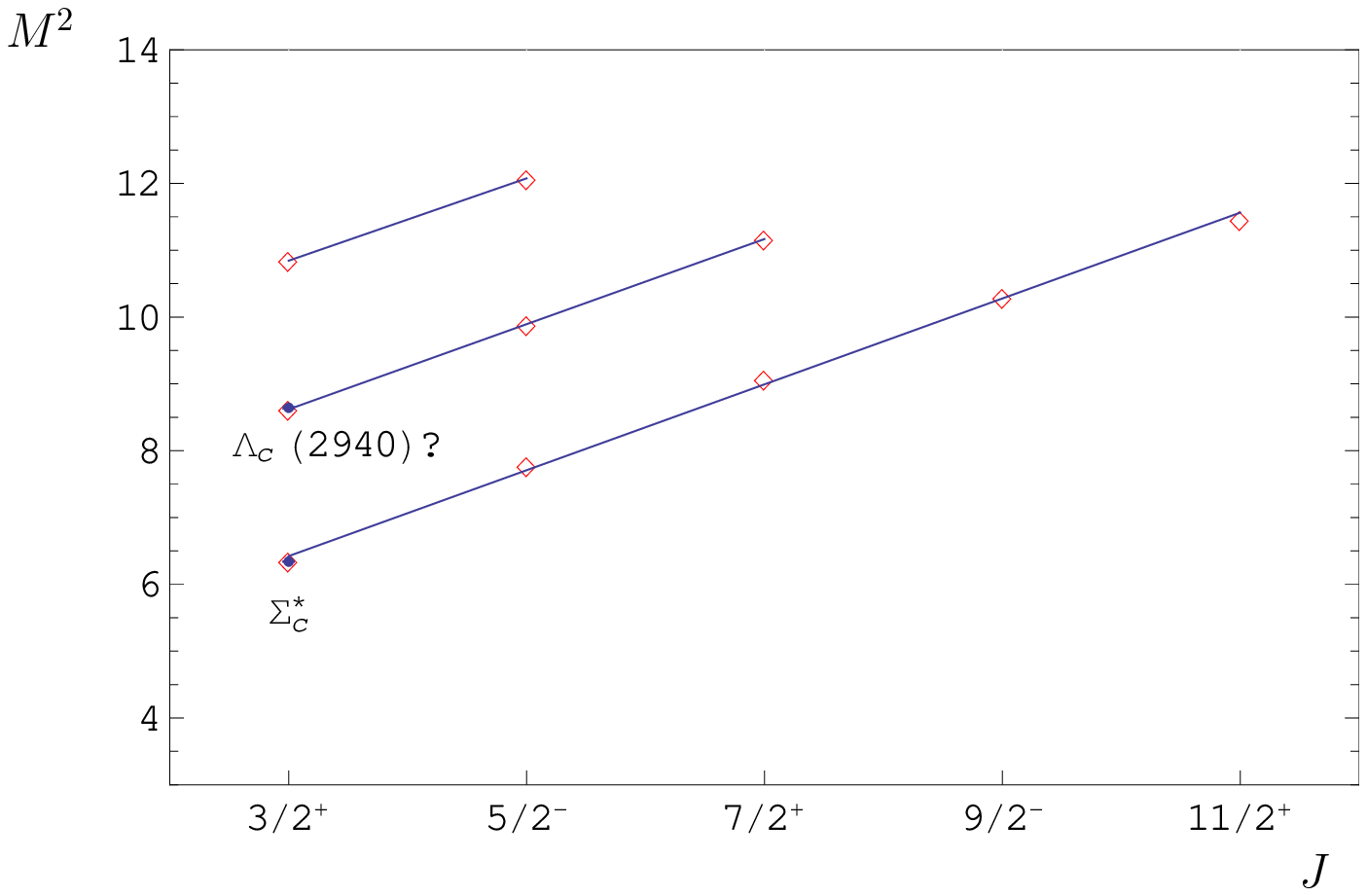}\\
\vspace*{-0.3cm}\hspace*{0.5cm} (a) \hspace*{7.5cm} (b) 

\caption{\label{fig:sigma_j} Same as in Fig.~\ref{fig:lambda_j} for
  the $\Sigma_c$ baryons. }\vspace*{0.5cm}
\end{figure}

\begin{figure}
 \includegraphics[width=8cm]{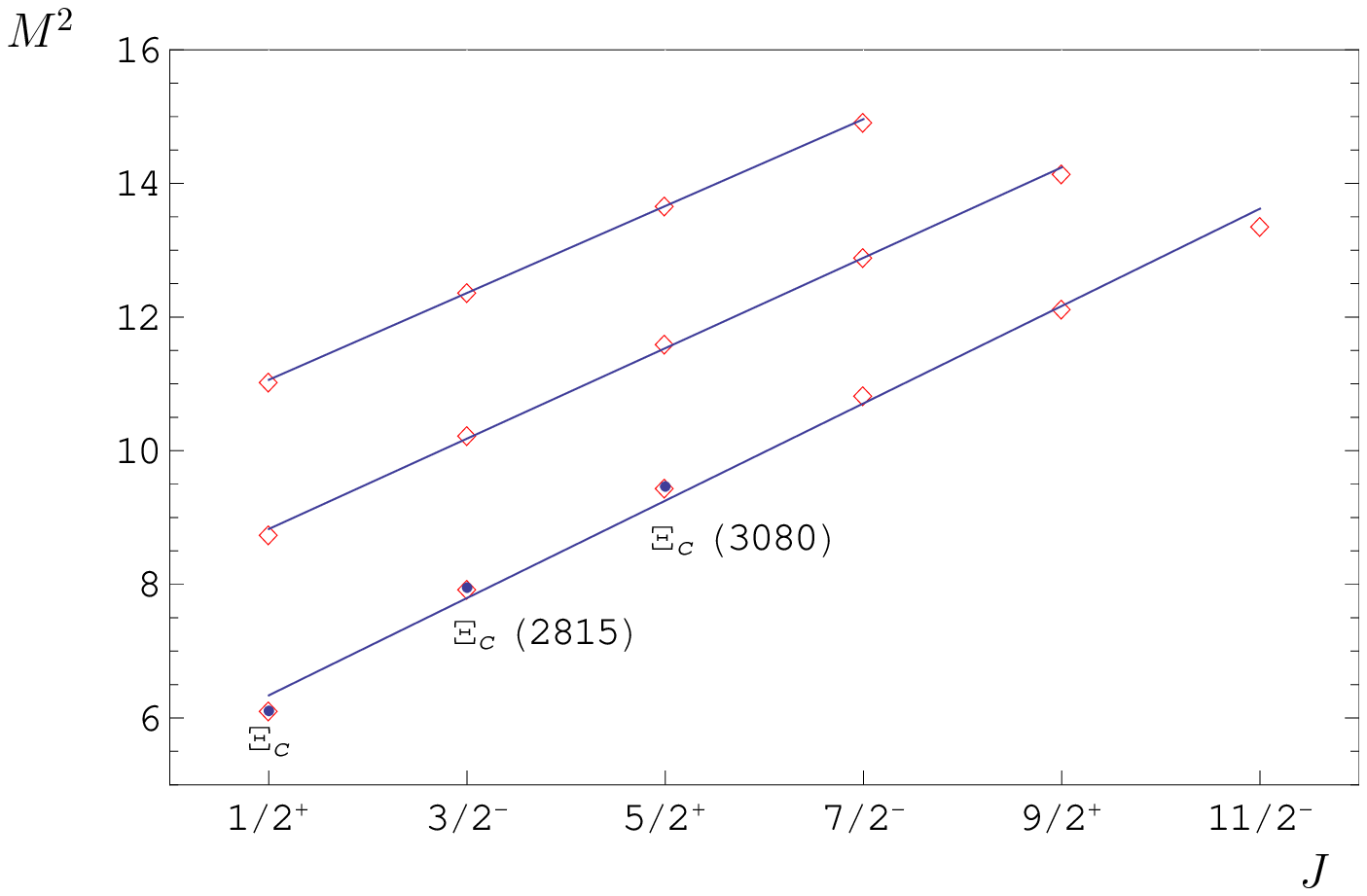}\ \ \ \includegraphics[width=8cm]{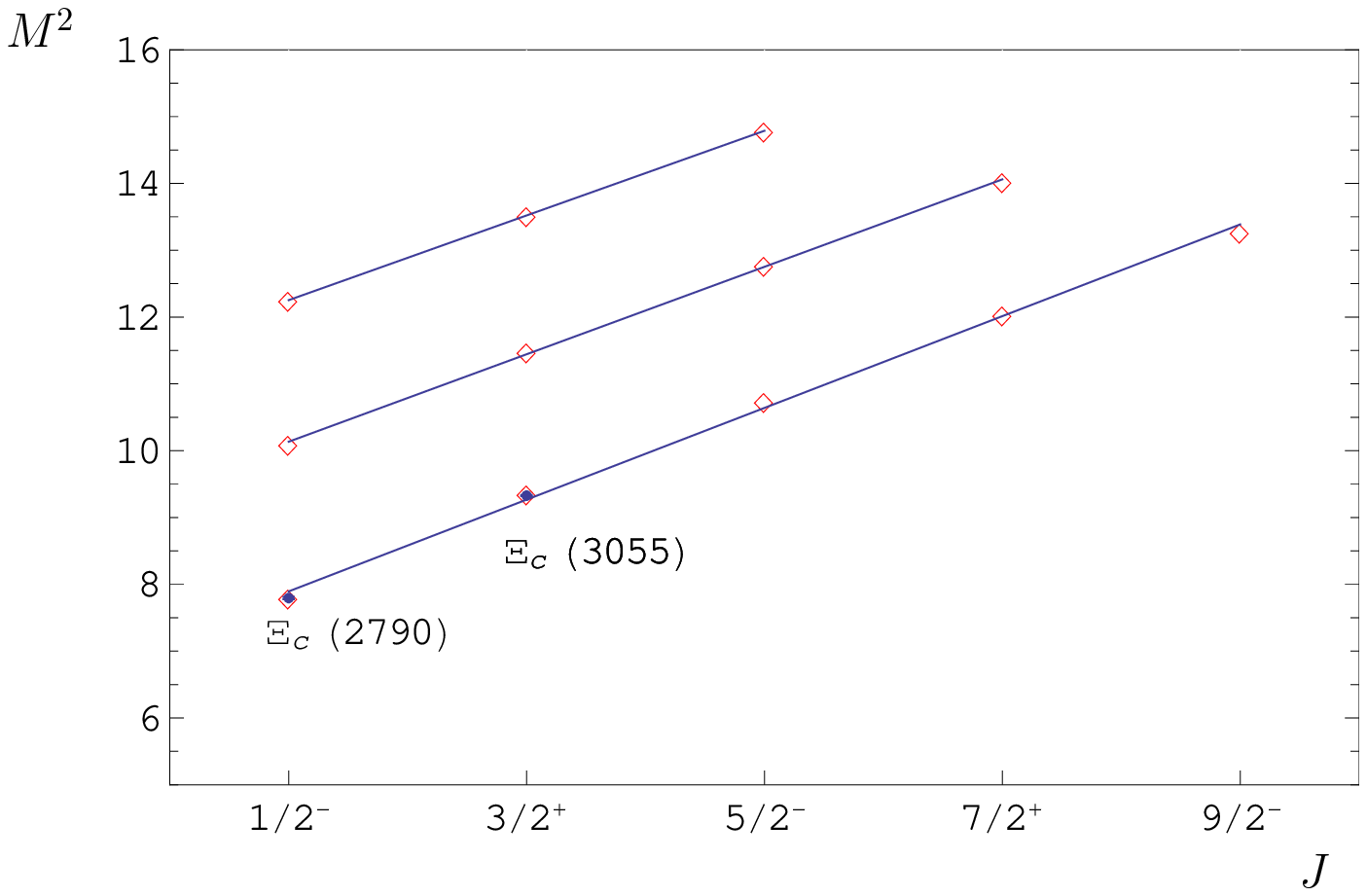}\\
\vspace*{-0.3cm}\hspace*{0.5cm} (a) \hspace*{7.5cm} (b) 

\caption{\label{fig:xi_sd_j} Same as in Fig.~\ref{fig:lambda_j} for
  the $\Xi_c$ baryons with the scalar diquark. } 
\end{figure}

\begin{figure}
 \includegraphics[width=8cm]{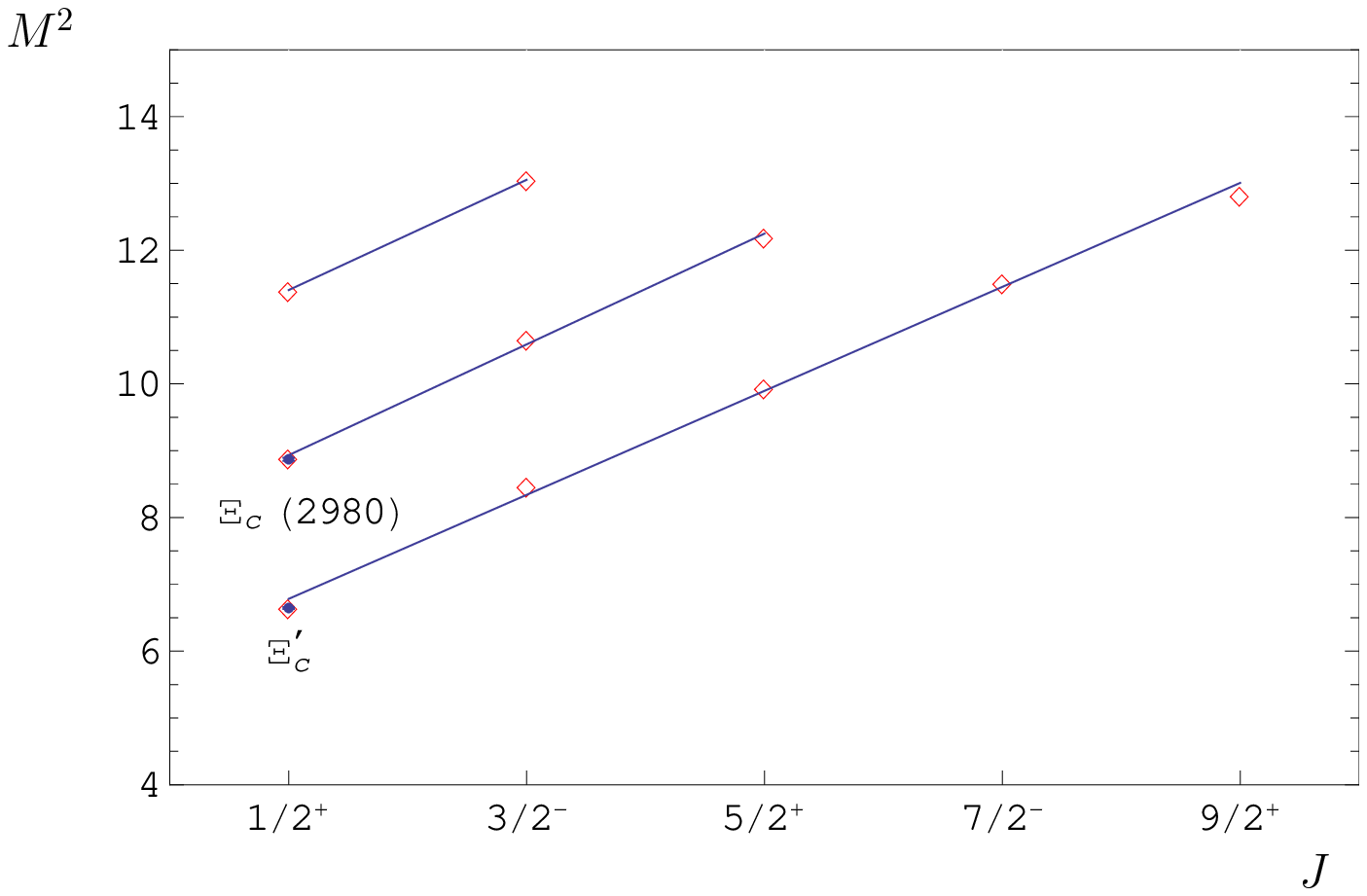}\ \ \ \includegraphics[width=8cm]{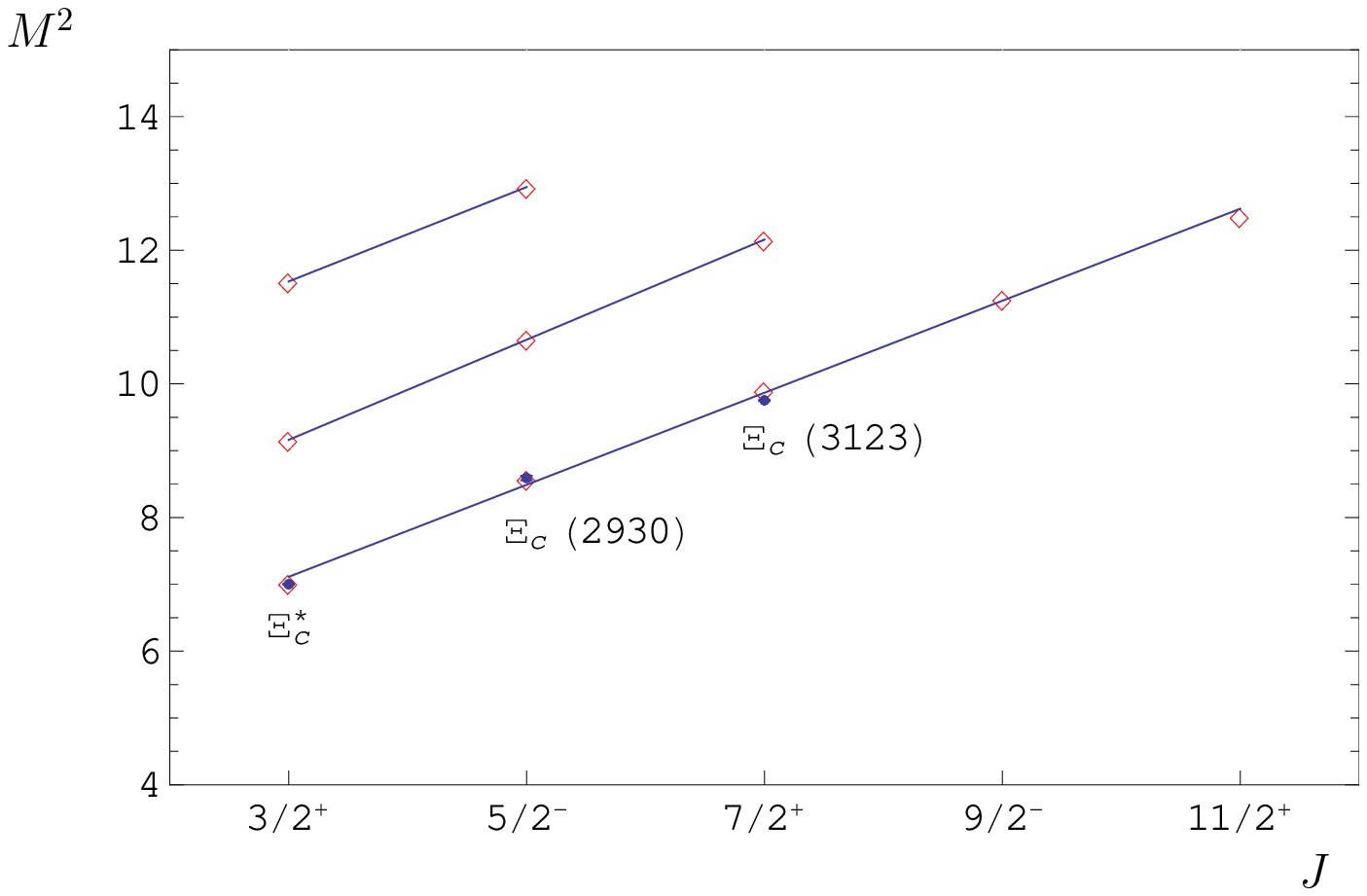}\\
\vspace*{-0.3cm}\hspace*{0.5cm} (a) \hspace*{7.5cm} (b) 

\caption{\label{fig:xi_vd_j} Same as in Fig.~\ref{fig:lambda_j} for
  the $\Xi_c'$ baryons with the axial vector diquark. }%\vspace*{0.5cm}
\end{figure}

\begin{figure}
 \includegraphics[width=8cm]{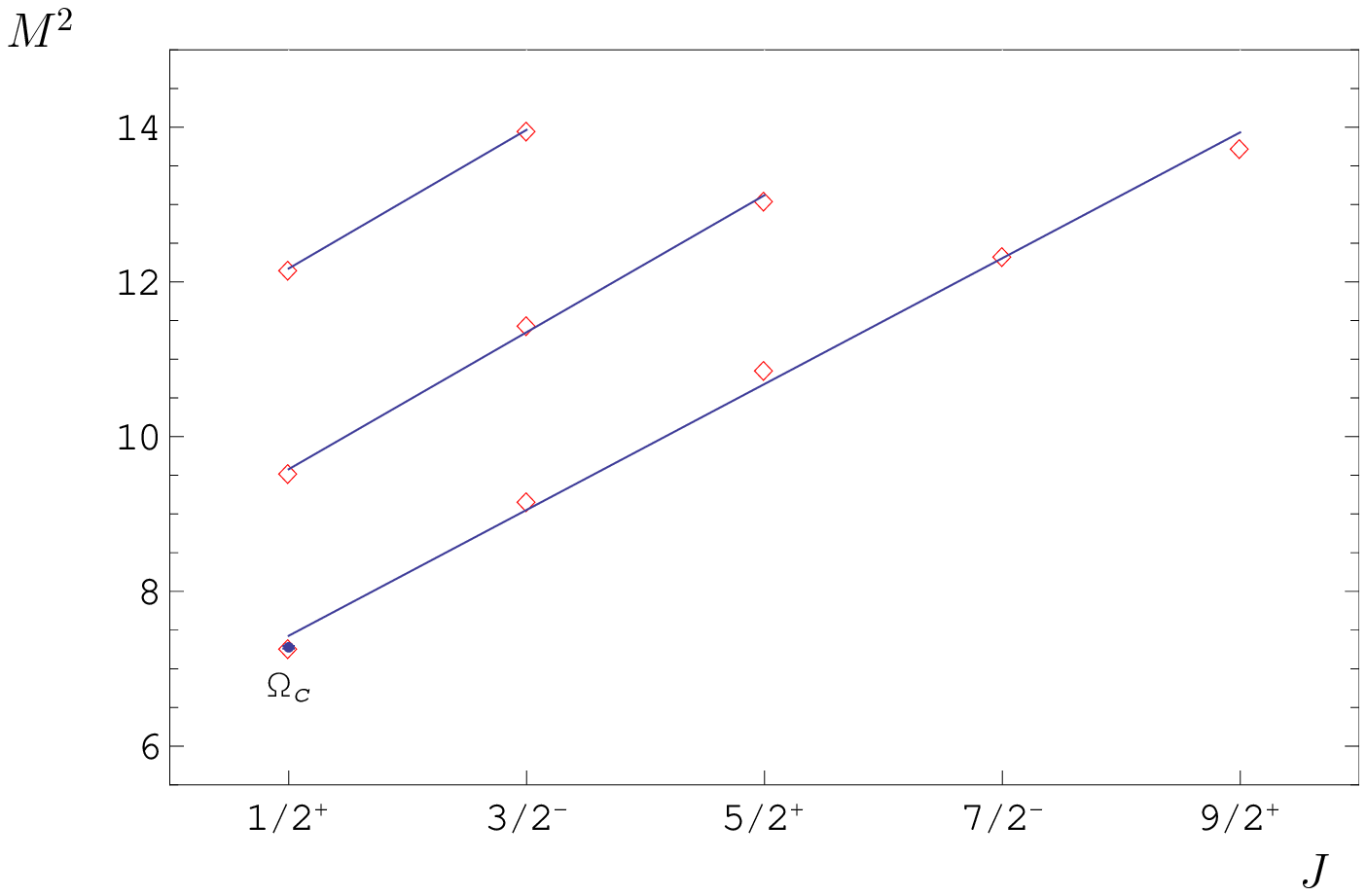}\ \ \ \includegraphics[width=8cm]{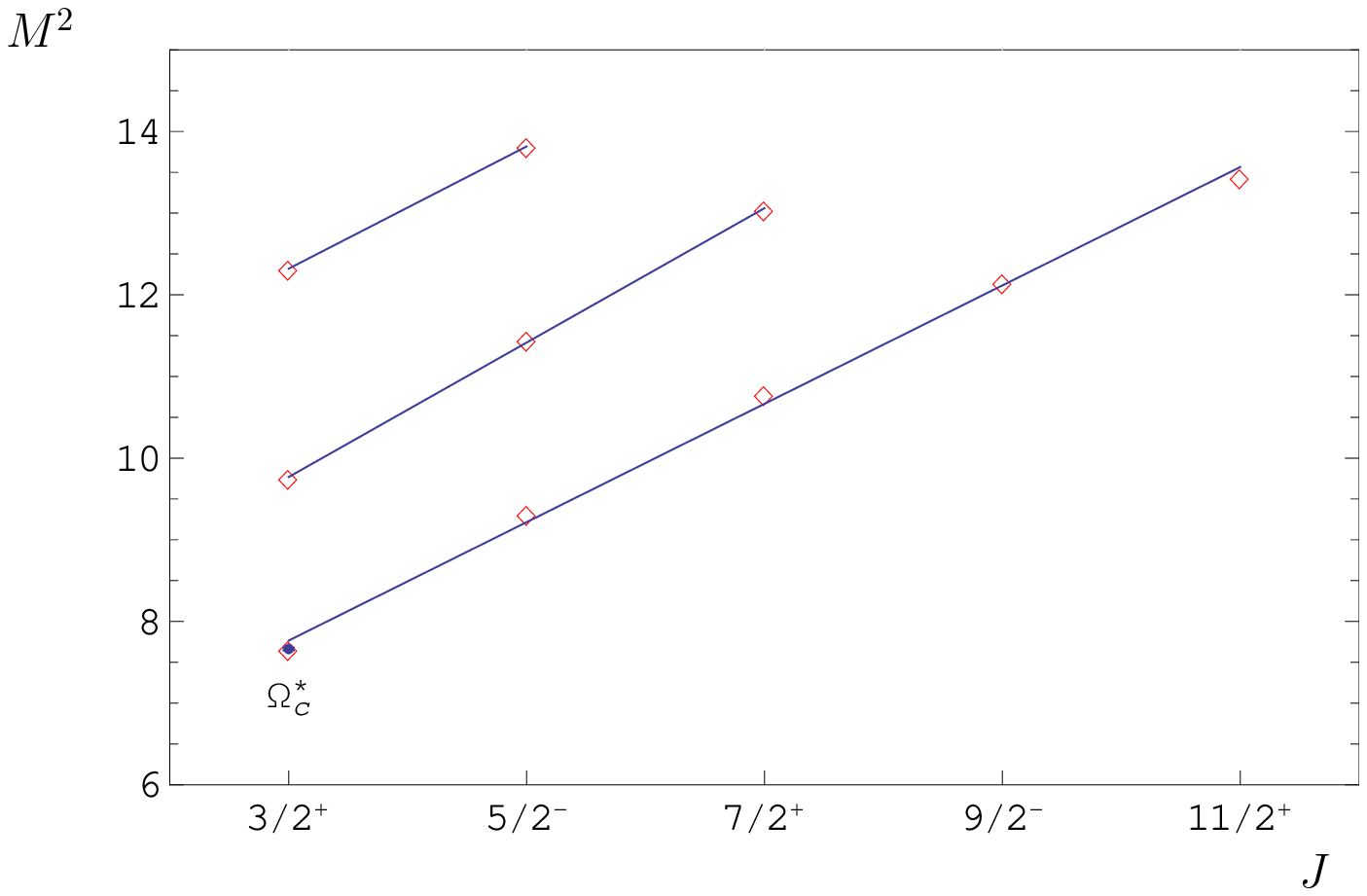}\\
\vspace*{-0.3cm}\hspace*{0.5cm} (a) \hspace*{7.5cm} (b) 

\caption{\label{fig:omega_j} Same as in Fig.~\ref{fig:lambda_j} for
  the $\Omega_c$ baryons. }
\end{figure}

\begin{figure}
  \includegraphics[width=8cm]{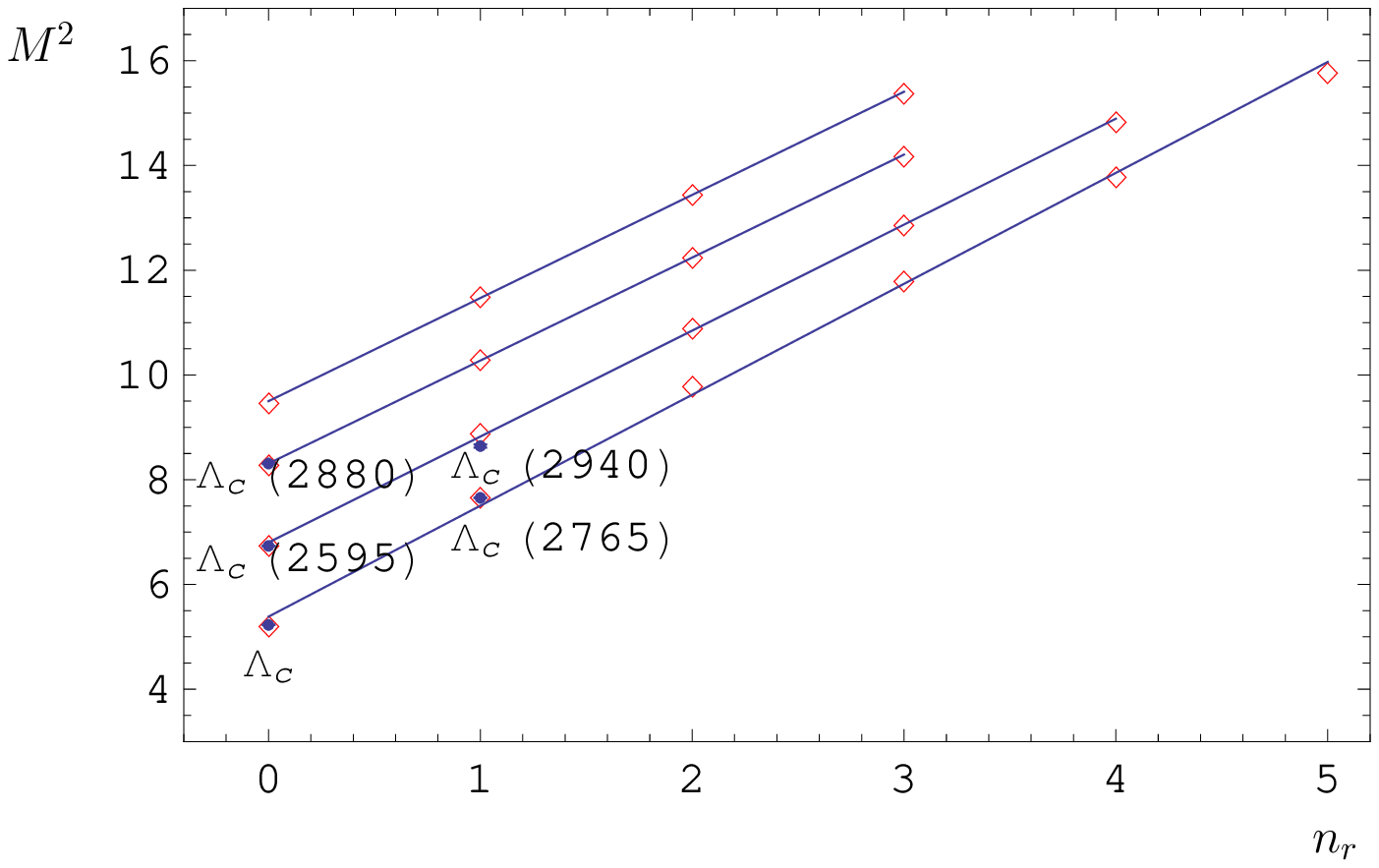}\ \ \  \includegraphics[width=8cm]{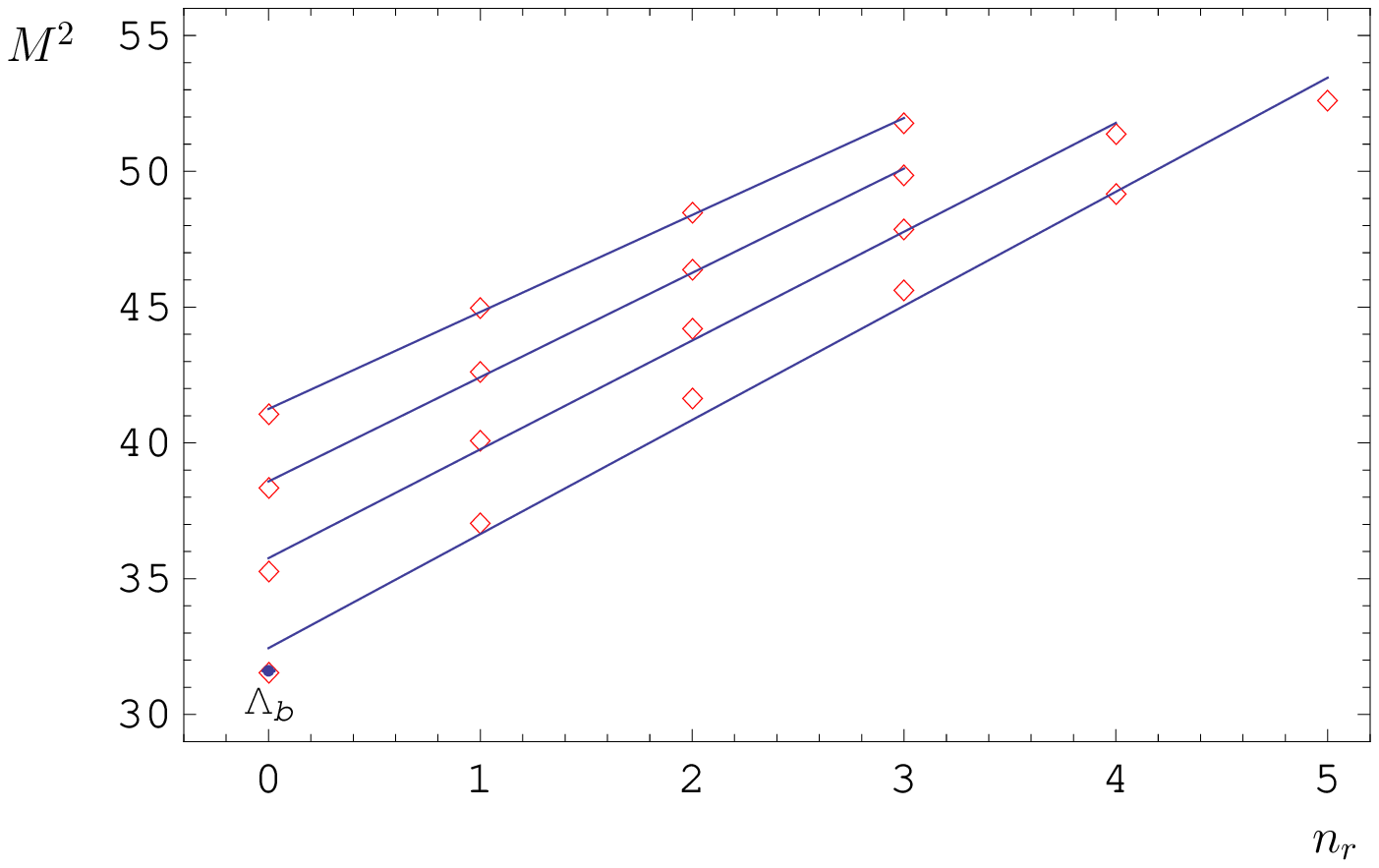}

\caption{\label{fig:lambda_n} The $(n_r,M^2)$ Regge trajectories for
  $\Lambda_Q\left(\frac12^+\right)$,
  $\Lambda_Q\left(\frac12^-\right)$,
  $\Lambda_Q\left(\frac52^+\right)$ and
  $\Lambda_Q\left(\frac72^+\right)$ baryons (from bottom to
  top). Notations are the same as in Fig.~\ref{fig:lambda_j}. }
\end{figure}

\begin{figure}
  \includegraphics[width=8cm]{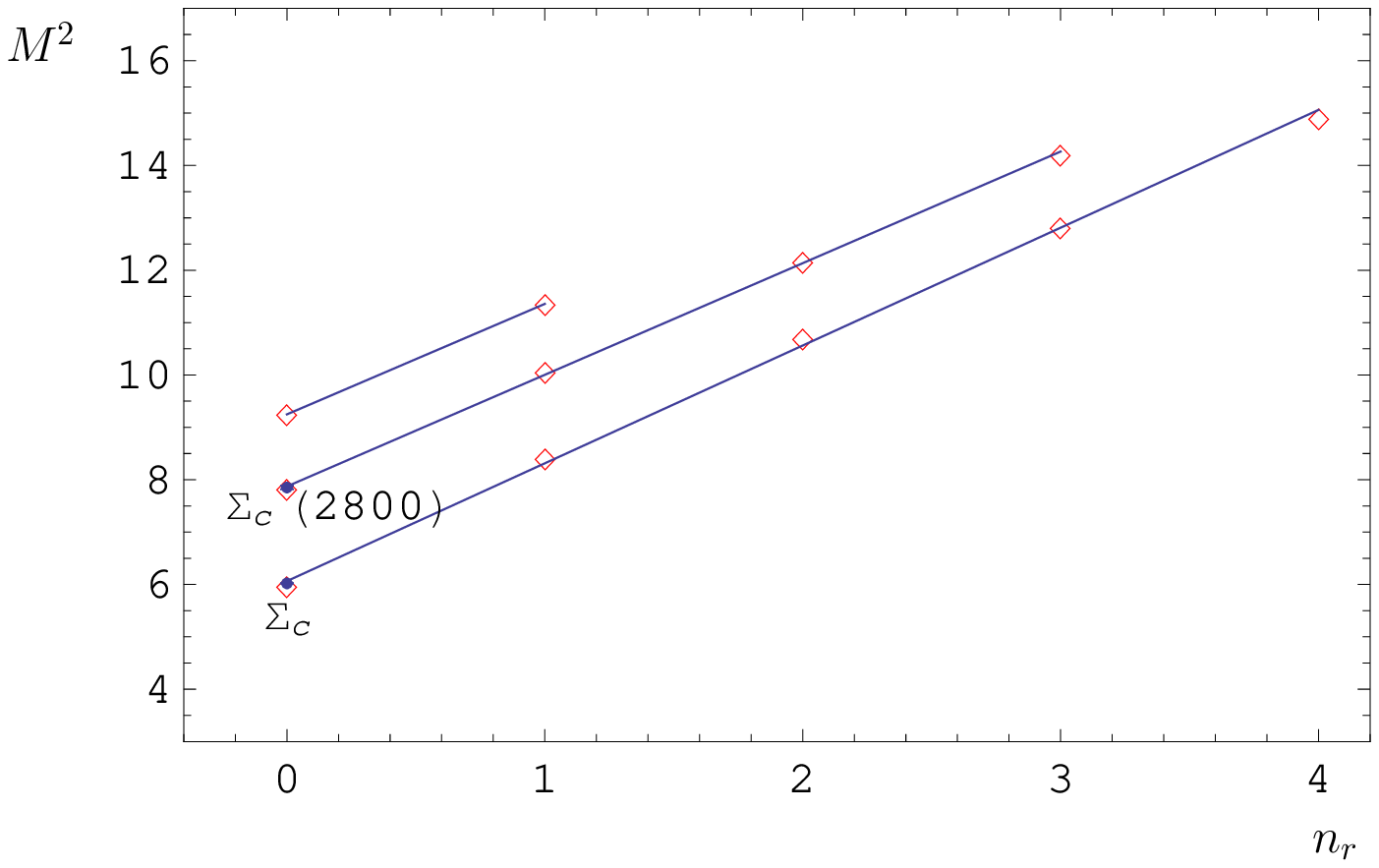}\ \ \  \includegraphics[width=8cm]{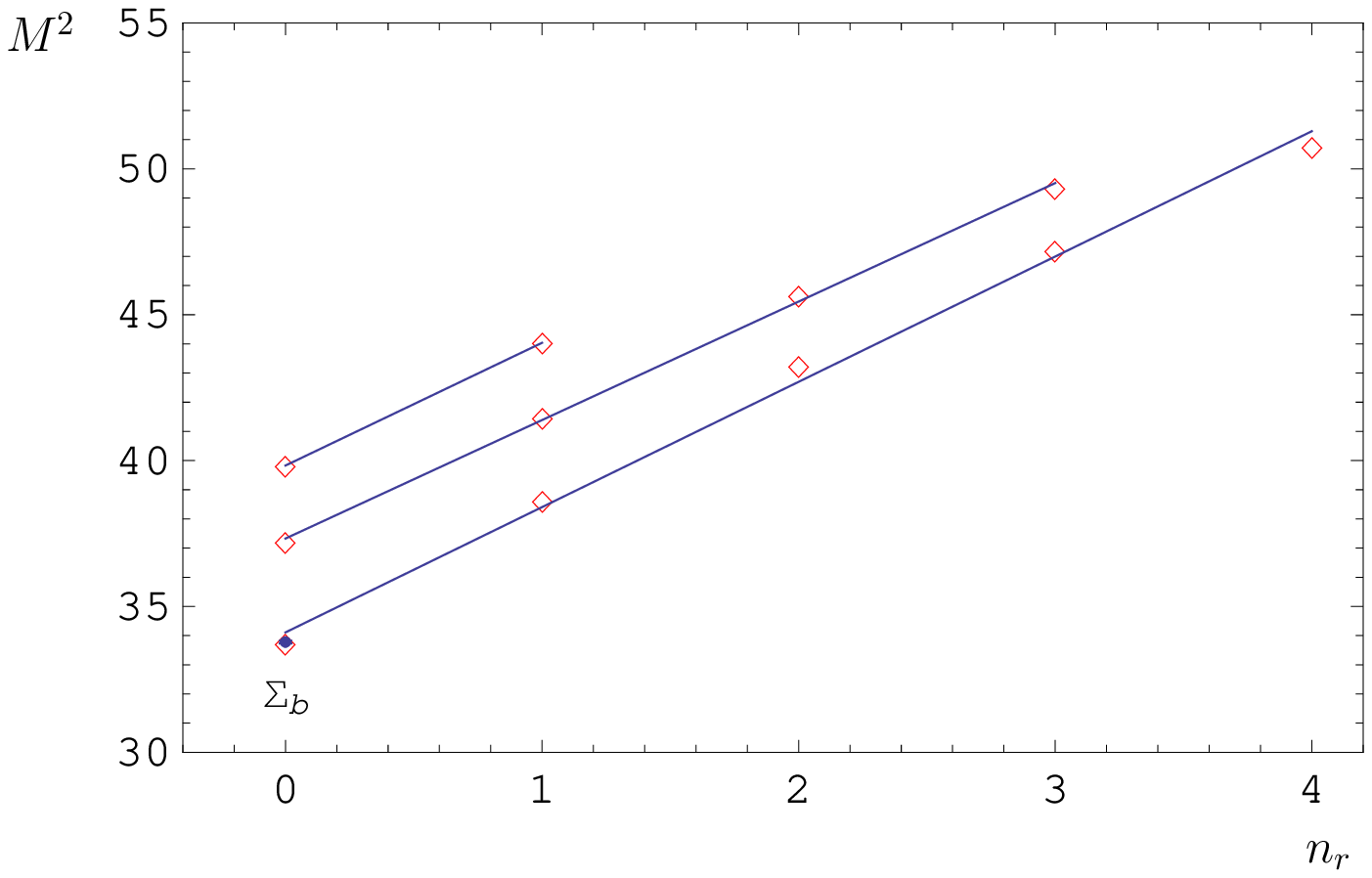}

\caption{\label{fig:sigma_n} The $(n_r,M^2)$ Regge trajectories for
  $\Sigma_Q\left(\frac12^+,S\right)$,
  $\Sigma_Q\left(\frac12^-,P\right)$ and
  $\Sigma_Q\left(\frac12^+,D\right)$ baryons (from bottom to
  top). Notations are the same as in Fig.~\ref{fig:lambda_j}. }
\end{figure}

\begin{figure}
  \includegraphics[width=8cm]{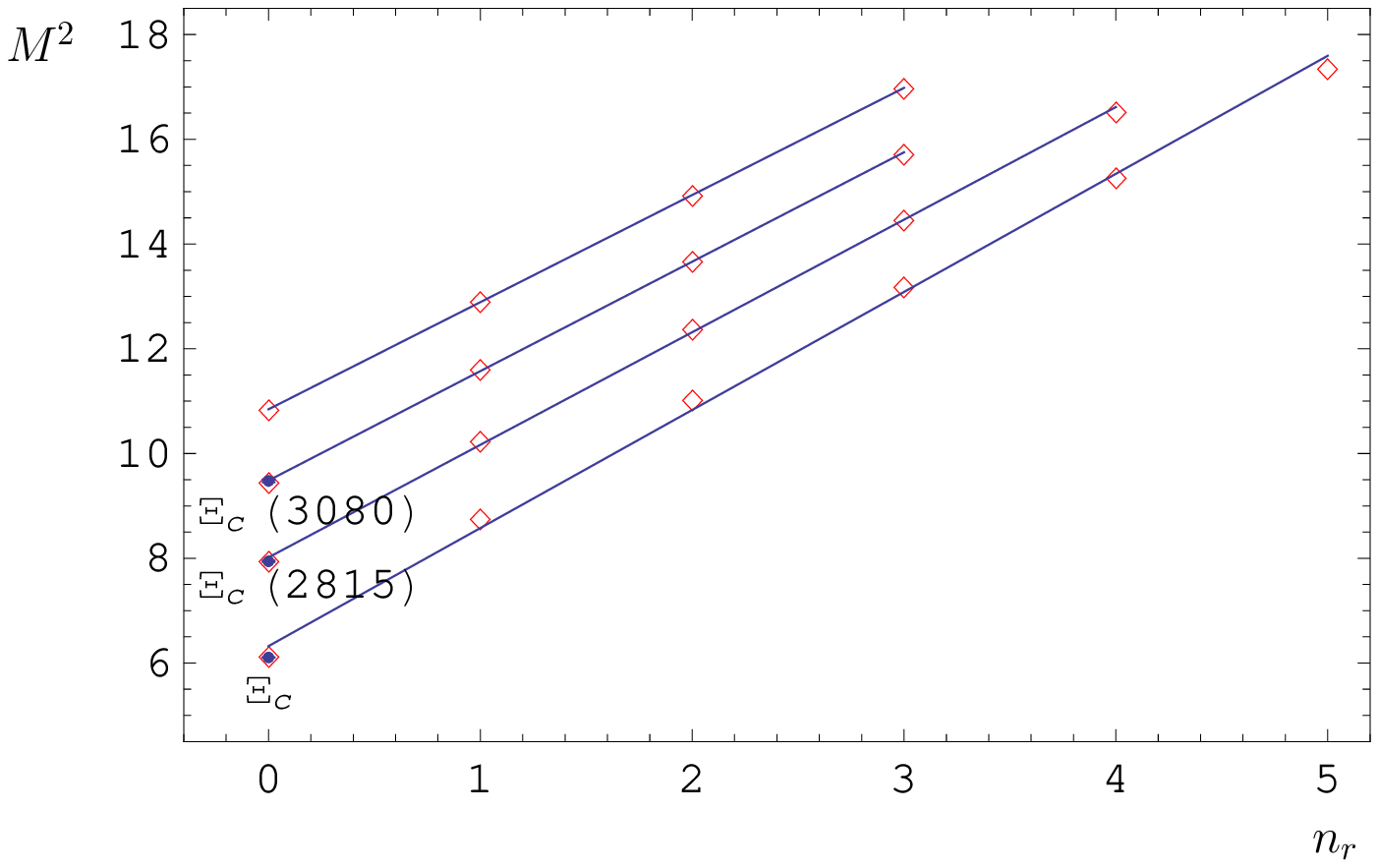}\ \ \  \includegraphics[width=8cm]{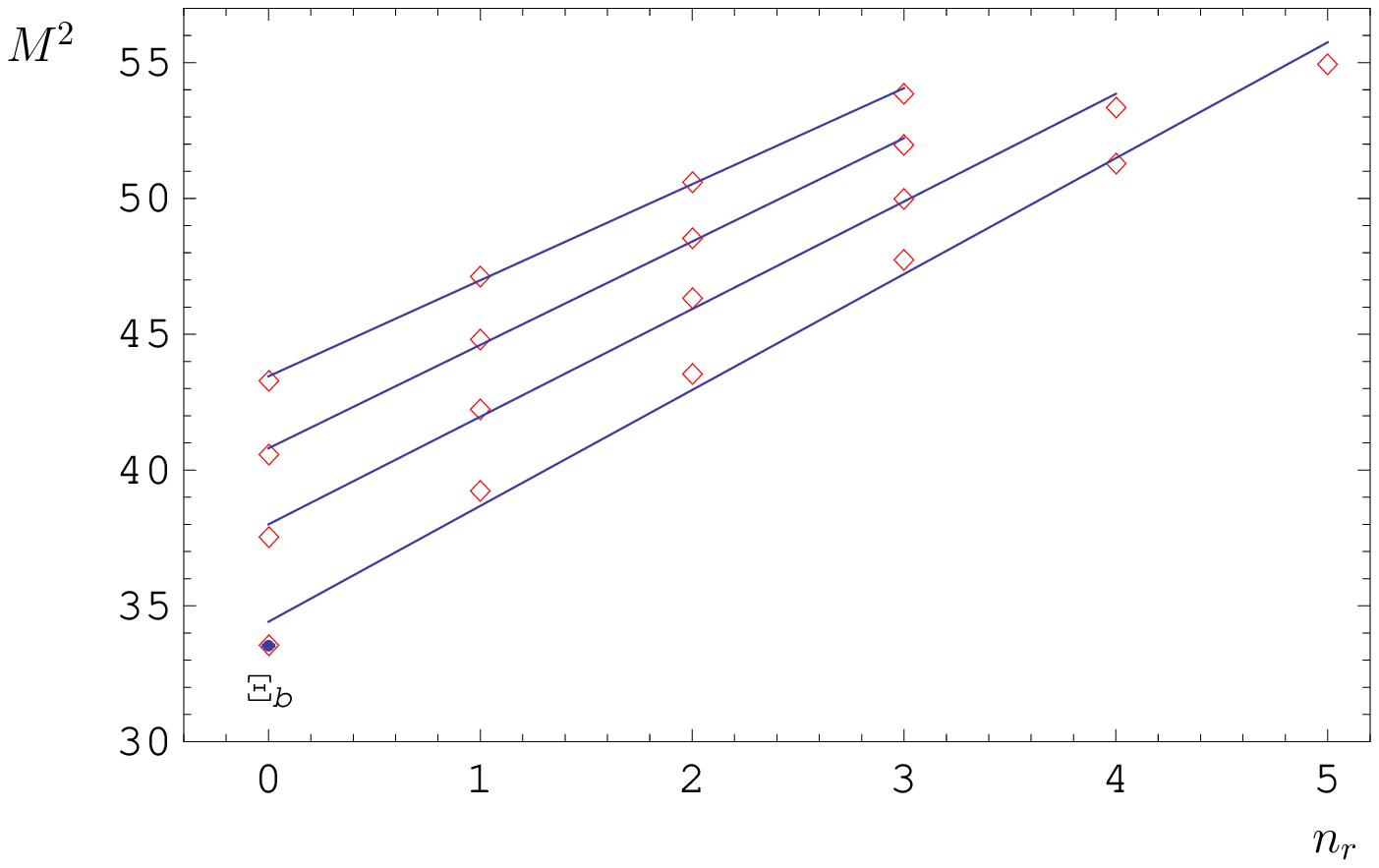}

\caption{\label{fig:xi_sd_n} The $(n_r,M^2)$ Regge trajectories for
  $\Xi_Q\left(\frac12^+\right)$,
  $\Xi_Q\left(\frac32^-\right)$,
  $\Xi_Q\left(\frac52^+\right)$ and
  $\Xi_Q\left(\frac72^-\right)$ baryons (from bottom to
  top) with the scalar diquark. Notations are the same as in Fig.~\ref{fig:lambda_j}. }
\end{figure}

\begin{figure}
  \includegraphics[width=8cm]{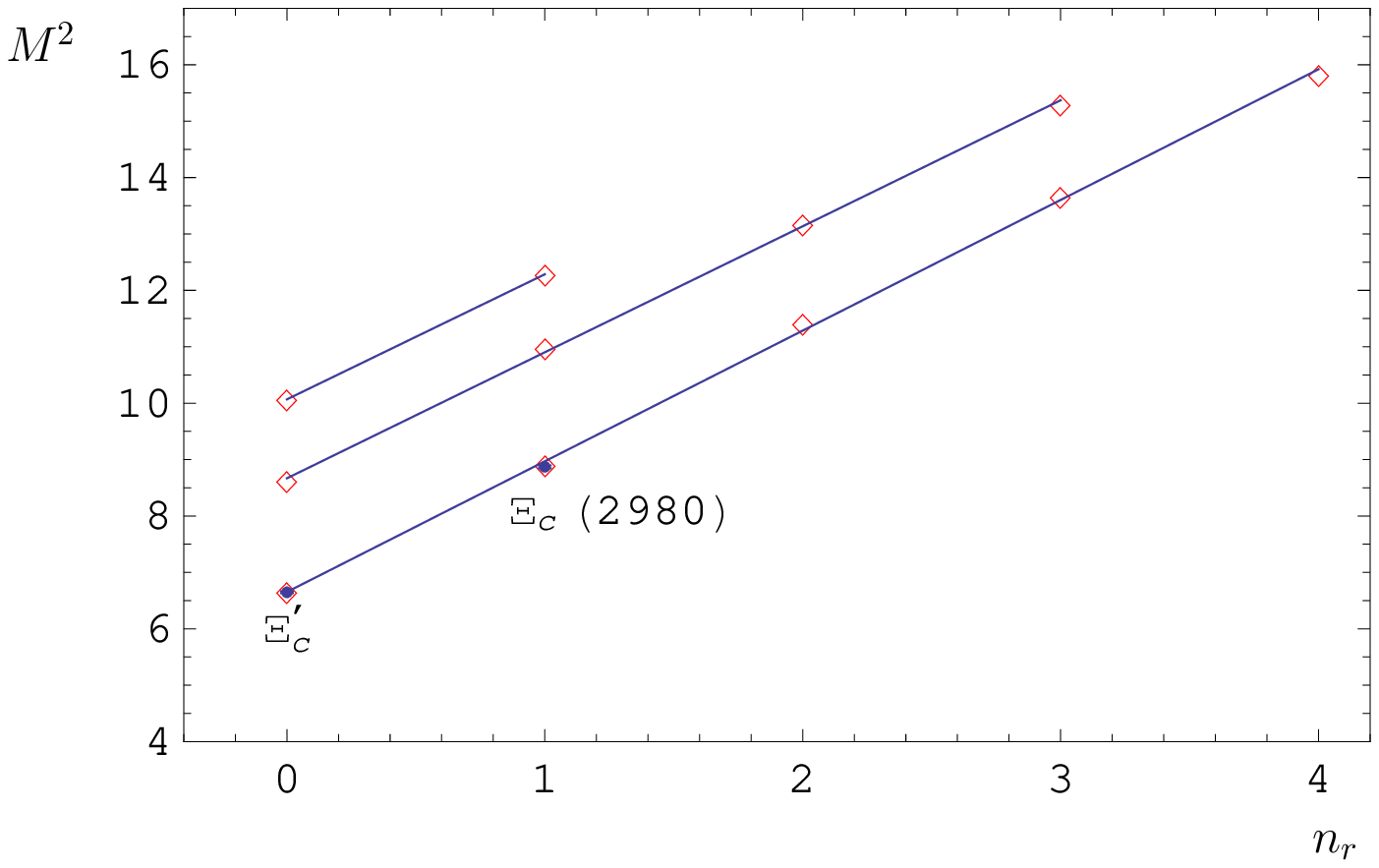}\ \ \  \includegraphics[width=8cm]{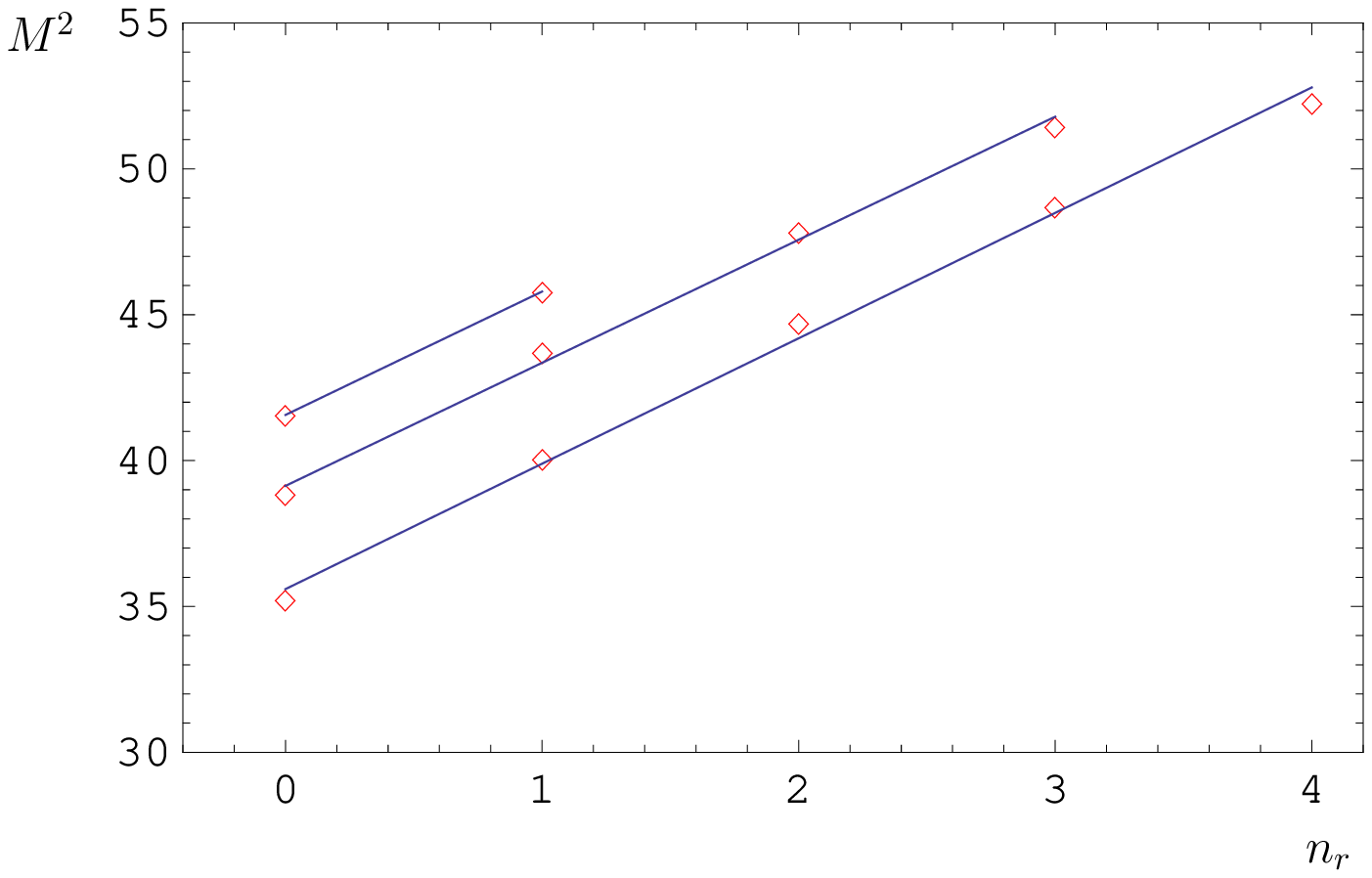}

\caption{\label{fig:xi_vd_n} The $(n_r,M^2)$ Regge trajectories for
  $\Xi'_Q\left(\frac12^+\right)$,
  $\Xi_Q\left(\frac12^-\right)$ and
  $\Xi_Q\left(\frac12^+\right)$ baryons (from bottom to
  top) with the axial vector diquark. Notations are the same as in Fig.~\ref{fig:lambda_j}. }
\end{figure}

\begin{figure}
  \includegraphics[width=8cm]{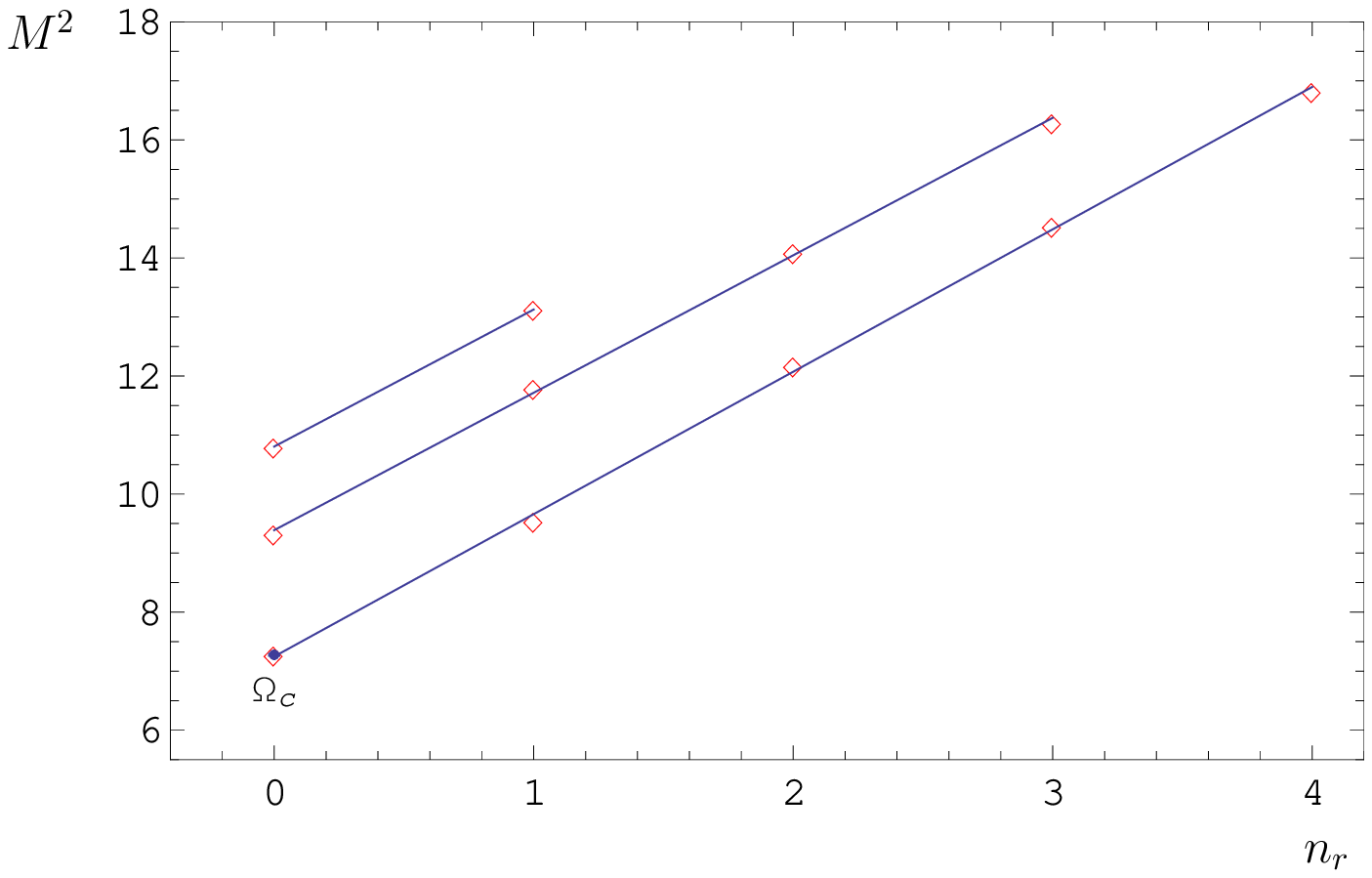}\ \ \  \includegraphics[width=8cm]{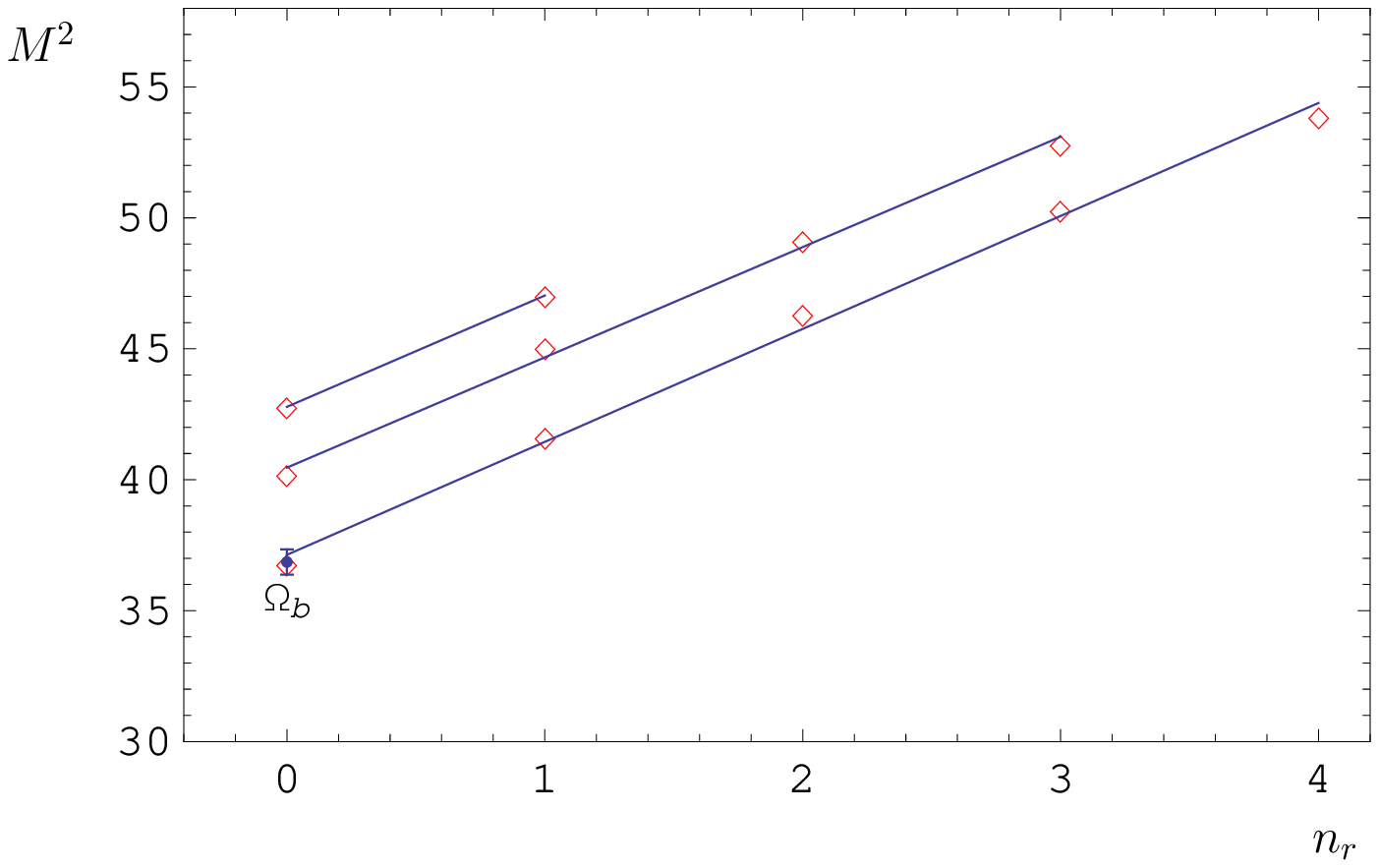}

\caption{\label{fig:omega_n} The $(n_r,M^2)$ Regge trajectories for
  $\Omega_Q\left(\frac12^+,S\right)$,
  $\Omega_Q\left(\frac12^-,P\right)$ and
  $\Omega_Q\left(\frac12^+,D\right)$ baryons (from bottom to
  top). Notations are the same as in Fig.~\ref{fig:lambda_j}. }
\end{figure}

In Figs.~\ref{fig:lambda_j}-\ref{fig:omega_j} we plot the Regge trajectories in
the ($J, M^2$) plane for charmed and bottom baryons with natural
($P=(-1)^{J-1/2}$) and unnatural ($P=(-1)^{J+1/2}$) parities \cite{col}. The Regge trajectories in the
$(n_r,M^2)$ plane are presented 
in Figs.~\ref{fig:lambda_n}-\ref{fig:omega_n}. The masses calculated in our
model are shown by diamonds. Available experimental data are given by
dots with error bars and corresponding baryon names. 
Straight lines were obtained by a
$\chi^2$ fit of the calculated values. The fitted slopes
and intercepts of the Regge trajectories are given in
Tables~\ref{tab:rtj} and \ref{tab:rtn}. We see that the calculated
heavy baryon masses fit nicely to the linear trajectories in both
planes. These trajectories are almost parallel and equidistant.

%\newpage

\begin{longtable}[t!]{@{ \ \  \ }c@{ \ \  \ \  \ \  \ \  \ \  \ \ }c@{ \ \  \ \  \ \  \ \  \ \  \ \  \ }c@{\ \ \ \ \  \ \  \ \  \ \  \ \  \ \  \ \  \ }c@{\ \ \ \ \  \ \  \ \  \ \  \ \  \ \  \ \  \ \  \ }c@{ \ \  \ \ }}
\caption{ Masses of the $\Xi_Q$ ($Q=c,b$) heavy baryons
  with the axial vector diquark (in MeV).}
\label{tab:xv}\vspace*{-0.5cm}\\
\hline
\hline
\\[1pt]
% \begin{ruledtabular}
% \begin{tabular}{ccccccc}
& & \multicolumn{2}{c}{\hspace{-2.2cm}\underline{\hspace{2.4cm}$Q=c$\hspace{2.4cm}}}\hspace{-0.8cm}& {\underline{\hspace{0.2cm}$Q=b$\hspace{0.2cm}}}\\
$I(J^P)$& $Qd$ state & $M$ & $M^{\rm exp}$  \cite{pdg}&  $M$ \\
\hline
\endfirsthead
\caption[]{(continued)}\vspace*{-0.5cm}\\
\hline\hline\\[1pt]
& & \multicolumn{2}{c}{\hspace{-2.2cm}\underline{\hspace{2.4cm}$Q=c$\hspace{2.4cm}}}\hspace{-0.8cm}& {\underline{\hspace{0.2cm}$Q=b$\hspace{0.2cm}}}\\
$I(J^P)$& $Qd$ state & $M$ & $M^{\rm exp}$  \cite{pdg}&  $M$ \\[2pt]
\hline
\endhead
\\[2pt]\hline
\hline
\endfoot
\endlastfoot
$\frac12(\frac12^+)$& $1S$ & 2579 & 2577.9(2.9) & 5936 \\
$\frac12(\frac12^+)$& $2S$ & 2983 & 2971.4(3.3) & 6329 \\
$\frac12(\frac12^+)$& $3S$ & 3377 & &             6687 \\
$\frac12(\frac12^+)$& $4S$ & 3695 & &             6978 \\
$\frac12(\frac12^+)$& $5S$ & 3978 & &             7229 \\
$\frac12(\frac32^+)$& $1S$ & 2649 & 2645.9(0.5) & 5963 \\
$\frac12(\frac32^+)$& $2S$ & 3026 &             & 6342 \\
$\frac12(\frac32^+)$& $3S$ & 3396 &             & 6695 \\
$\frac12(\frac32^+)$& $4S$ & 3709 &             & 6984 \\
$\frac12(\frac32^+)$& $5S$ & 3989 &             & 7234 \\
$\frac12(\frac12^-)$& $1P$ & 2936 &   2931(6)          & 6233    \\
$\frac12(\frac12^-)$& $2P$ & 3313 &             & 6611 \\
$\frac12(\frac12^-)$& $3P$ & 3630 &             & 6915 \\
$\frac12(\frac12^-)$& $4P$ & 3912 &             & 7174 \\
$\frac12(\frac12^-)$& $1P$ & 2854 &             & 6227     \\
$\frac12(\frac12^-)$& $2P$ & 3267 &            & 6604 \\
$\frac12(\frac12^-)$& $3P$ & 3598 &            & 6906 \\
$\frac12(\frac12^-)$& $4P$ & 3887 &             & 7164 \\
$\frac12(\frac32^-)$& $1P$ & 2935 &  2931(6)          & 6234  \\
$\frac12(\frac32^-)$& $2P$ & 3311 &             & 6605 \\
$\frac12(\frac32^-)$& $3P$ & 3628 &             & 6905 \\
$\frac12(\frac32^-)$& $4P$ & 3911 &            & 7163 \\
$\frac12(\frac32^-)$& $1P$ & 2912 &             & 6224  \\
$\frac12(\frac32^-)$& $2P$ & 3293 &             & 6598  \\
$\frac12(\frac32^-)$& $3P$ & 3613 &             & 6900  \\
$\frac12(\frac32^-)$& $4P$ & 3898 &            & 7159  \\
$\frac12(\frac52^-)$& $1P$ & 2929 & 2931(6)    & 6226  \\
$\frac12(\frac52^-)$& $2P$ & 3303 &             & 6596\\
$\frac12(\frac52^-)$& $3P$ & 3619 &            & 6897\\
$\frac12(\frac52^-)$& $4P$ & 3902 &            & 7156\\
$\frac12(\frac12^+)$& $1D$ & 3163 &             & 6447  \\
$\frac12(\frac12^+)$& $2D$ & 3505 &             & 6767  \\
$\frac12(\frac32^+)$& $1D$ & 3167 &             & 6459 \\
$\frac12(\frac32^+)$& $2D$ & 3506 &             & 6775 \\
$\frac12(\frac32^+)$& $1D$ & 3160 &            & 6431 \\
$\frac12(\frac32^+)$& $2D$ & 3497 &             & 6751 \\
$\frac12(\frac52^+)$& $1D$ & 3166 &             & 6432 \\
$\frac12(\frac52^+)$& $2D$ & 3504 &            & 6751 \\
$\frac12(\frac52^+)$& $1D$ & 3153 &            & 6420 \\
$\frac12(\frac52^+)$& $2D$ & 3493 &             & 6740 \\
$\frac12(\frac72^+)$& $1D$ & 3147 & 3122.9(1.3)   & 6414 \\
$\frac12(\frac72^+)$& $2D$ & 3486 &              & 6736 \\
$\frac12(\frac32^-)$& $1F$ & 3418 &             & 6675  \\
$\frac12(\frac52^-)$& $1F$ & 3408 &             & 6686  \\
$\frac12(\frac52^-)$& $1F$ & 3394 &             & 6640  \\
$\frac12(\frac72^-)$& $1F$ & 3393 &              & 6641  \\
$\frac12(\frac72^-)$& $1F$ & 3373 &              & 6619  \\
$\frac12(\frac92^-)$& $1F$ & 3357 &             & 6610  \\
$\frac12(\frac52^+)$& $1G$ & 3623 &              & 6867  \\
$\frac12(\frac72^+)$& $1G$ & 3608 &              & 6876  \\
$\frac12(\frac72^+)$& $1G$ & 3584 &              & 6822  \\
$\frac12(\frac92^+)$& $1G$ & 3582 &             & 6821  \\
$\frac12(\frac92^+)$& $1G$ & 3558 &              & 6792  \\
$\frac12(\frac{11}2^+)$& $1G$ &3536 &            & 6782 \\[2pt]
\hline\hline
% \end{tabular}
% \end{ruledtabular}
\end{longtable}

\begin{longtable}{@{ \ }c@{\  \ \  \ \  \ \ }c@{\ \ \ \  \ \  \ \  \ \
    }c@{\ \ \ \  \ \  \ \  \ \  \ \  \ \ }c@{\ \ \ \ \ \  \ \  \ \ \ \
      \  \ \ }c@{\ \ \ \ \ \  \ \  \ \  \ \  \ \ }c@{ \ }}
\caption{ Masses of the $\Omega_Q$ ($Q=c,b$) heavy baryons (in MeV).}
% \begin{ruledtabular}
% \begin{tabular}{cccccc}
\label{tab:om}\vspace*{-0.5cm}\\
\hline
\hline
\\[1pt]
& & \multicolumn{2}{c}{\hspace{-1.5cm}\underline{\hspace{2.1cm}$Q=c$\hspace{2.1cm}}}\hspace{-0.8cm}& \multicolumn{2}{c}{\hspace{-0.4cm}\underline{\hspace{2.cm}$Q=b$\hspace{2.cm}}}\\
$I(J^P)$& $Qd$ state & $M$ & $M^{\rm exp}$  \cite{pdg} &  $M$& $M^{\rm exp}$  \cite{pdg} \\
\hline
\endfirsthead
\caption[]{(continued)}\vspace*{-0.5cm}\\
\hline\hline\\[1pt]
& & \multicolumn{2}{c}{\hspace{-1.5cm}\underline{\hspace{2.1cm}$Q=c$\hspace{2.1cm}}}\hspace{-0.8cm}& \multicolumn{2}{c}{\hspace{-0.4cm}\underline{\hspace{2.cm}$Q=b$\hspace{2.cm}}}\\
$I(J^P)$& $Qd$ state & $M$ & $M^{\rm exp}$  \cite{pdg} &  $M$ & $M^{\rm exp}$  \cite{pdg} \\[2pt]
\hline
\endhead
\\[2pt]\hline
\hline
\endfoot
\endlastfoot
$0(\frac12^+)$& $1S$ & 2698 & 2695.2(1.7) & 6064& 6071(40)\\
$0(\frac12^+)$& $2S$ & 3088 &             & 6450 \\
$0(\frac12^+)$& $3S$ & 3489 &             & 6804 \\
$0(\frac12^+)$& $4S$ & 3814 &             & 7091 \\
$0(\frac12^+)$& $5S$ & 4102 &             & 7338 \\
$0(\frac32^+)$& $1S$ & 2768 & 2765.9(2.0) & 6088\\
$0(\frac32^+)$& $2S$ & 3123 &             & 6461 \\
$0(\frac32^+)$& $3S$ & 3510 &             & 6811 \\
$0(\frac32^+)$& $4S$ & 3830 &             & 7096 \\
$0(\frac32^+)$& $5S$ & 4114 &             & 7343 \\
$0(\frac12^-)$& $1P$ & 3055 &             & 6339     \\
$0(\frac12^-)$& $2P$ & 3435 &             & 6710 \\
$0(\frac12^-)$& $3P$ & 3754 &             & 7009 \\
$0(\frac12^-)$& $4P$ & 4037 &             & 7265 \\
$0(\frac12^-)$& $1P$ & 2966 &             & 6330   \\
$0(\frac12^-)$& $2P$ & 3384 &             & 6706 \\
$0(\frac12^-)$& $3P$ & 3717 &             & 7003 \\
$0(\frac12^-)$& $2P$ & 4009 &             & 7257 \\
$0(\frac32^-)$& $1P$ & 3054 &             & 6340 \\
$0(\frac32^-)$& $2P$ & 3433 &             & 6705 \\
$0(\frac32^-)$& $3P$ & 3752 &             & 7002 \\
$0(\frac32^-)$& $4P$ & 4036 &             & 7258 \\
$0(\frac32^-)$& $1P$ & 3029 &             & 6331\\
$0(\frac32^-)$& $2P$ & 3415 &             & 6699 \\
$0(\frac32^-)$& $3P$ & 3737 &             & 6998 \\
$0(\frac32^-)$& $4P$ & 4023 &             & 7250 \\
$0(\frac52^-)$& $1P$ & 3051 &             & 6334  \\
$0(\frac52^-)$& $2P$ & 3427 &             & 6700 \\
$0(\frac52^-)$& $3P$ & 3744 &             & 6996 \\
$0(\frac52^-)$& $4P$ & 4028 &             & 7251 \\
$0(\frac12^+)$& $1D$ & 3287 &             & 6540 \\
$0(\frac12^+)$& $2D$ & 3623 &             & 6857 \\
$0(\frac32^+)$& $1D$ & 3298 &             & 6549  \\
$0(\frac32^+)$& $2D$ & 3627 &             & 6863  \\
$0(\frac32^+)$& $1D$ & 3282 &             & 6530 \\
$0(\frac32^+)$& $2D$ & 3613 &             & 6846 \\
$0(\frac52^+)$& $1D$ & 3297 &             & 6529 \\
$0(\frac52^+)$& $2D$ & 3626 &             & 6846 \\
$0(\frac52^+)$& $1D$ & 3286 &             & 6520  \\
$0(\frac52^+)$& $2D$ & 3614 &             & 6837  \\
$0(\frac72^+)$& $1D$ & 3283 &             & 6517  \\
$0(\frac72^+)$& $2D$ & 3611 &             & 6834  \\
$0(\frac32^-)$& $1F$ & 3533 &              & 6763  \\
$0(\frac52^-)$& $1F$ & 3522 &             & 6771  \\
$0(\frac52^-)$& $1F$ & 3515 &             & 6737  \\
$0(\frac72^-)$& $1F$ & 3514 &             & 6736  \\
$0(\frac72^-)$& $1F$ & 3498 &             & 6719  \\
$0(\frac92^-)$& $1F$ & 3485 &             & 6713  \\
$0(\frac52^+)$& $1G$ & 3739 &             & 6952  \\
$0(\frac72^+)$& $1G$ & 3721 &            & 6959  \\
$0(\frac72^+)$& $1G$ & 3707 &            & 6916  \\
$0(\frac92^+)$& $1G$ & 3705 &            & 6915  \\
$0(\frac92^+)$& $1G$ & 3685 &            & 6892  \\
$0(\frac{11}2^+)$& $1G$ &3665 &           & 6884 \\[2pt]
\hline\hline
% \end{tabular}
% \end{ruledtabular}
\end{longtable}

\begin{table}
  \caption{Fitted parameters $\alpha$, $\alpha_0$ for the slope and intercept of the $(J,M^2)$ parent and daughter Regge
    trajectories for heavy baryons with scalar ($[q',q]$) and
    axial vector ($\{q',q\}$) diquark ($q=u,d$, $q'=u,d,s$).} 
  \label{tab:rtj}
\begin{ruledtabular}
\begin{tabular}{ccccc}
Trajectory&
$\alpha$ (GeV$^{-2}$)& $\alpha_0$&$\alpha$  (GeV$^{-2}$)&$\alpha_0$\\
\hline
$c[u,d]$ &$\Lambda_c\left(\frac12^+\right)$&&$\Lambda_c\left(\frac12^-\right)$\\
parent& $0.741\pm0.024$& $-3.504\pm0.205$& $0.782\pm0.030$&
$-4.874\pm0.276$\\
1 daughter &$0.793\pm0.013$&$-5.626\pm0.129$ & $0.815\pm0.009$&
$-6.769\pm0.099$\\
2 daughter &$0.821\pm0.005$&$-7.556\pm0.052$ & $0.839\pm0.004$&
$-8.654\pm0.043$\\
$c\{q,q\}$ &$\Sigma_c\left(\frac12^+\right)$&&$\Sigma_c^*\left(\frac32^+\right)$\\
parent& $0.679\pm0.032$& $-3.670\pm0.278$& $0.778\pm0.019$&
$-3.498\pm0.164$\\
1 daughter &$0.686\pm0.016$&$-5.289\pm0.158$ & $0.785\pm0.001$&
$-5.264\pm0.012$\\
2 daughter &$0.688$&$-6.865$ & $0.812$&
$-7.303$\\
$c[s,q]$ &$\Xi_c\left(\frac12^+\right)$&&$\Xi_c\left(\frac12^-\right)$\\
parent& $0.686\pm0.025$& $-3.852\pm0.240$& $0.728\pm0.020$&
$-5.249\pm0.211$\\
1 daughter &$0.739\pm0.015$&$-6.025\pm0.169$ & $0.764\pm0.012$&
$-7.244\pm0.142$\\
2 daughter &$0.769\pm0.008$&$-8.006\pm0.103$ & $0.789\pm0.004$&
$-9.168\pm0.052$\\
$c\{s,q\}$ &$\Xi_c'\left(\frac12^+\right)$&&$\Xi_c^*\left(\frac32^+\right)$\\
parent& $0.643\pm0.021$& $-3.856\pm0.212$& $0.726\pm0.019$&
$-3.665\pm0.191$\\
1 daughter &$0.603\pm0.026$&$-4.888\pm0.272$ & $0.667\pm0.005$&
$-4.614\pm0.051$\\
2 daughter &$0.606$&$-6.413$ & $0.708$&
$-6.865$\\
$c\{s,s\}$ &$\Omega_c\left(\frac12^+\right)$&&$\Omega_c^*\left(\frac32^+\right)$\\
parent& $0.615\pm0.023$& $-4.065\pm0.023$& $0.690\pm0.020$&
$-3.858\pm0.205$\\
1 daughter &$0.565\pm0.028$&$-4.910\pm0.316$ & $0.608\pm0.012$&
$-4.436\pm0.133$\\
2 daughter &$0.558$&$-6.293$ & $0.668$&
$-6.735$\\
$b[u,d]$ &$\Lambda_b\left(\frac12^+\right)$&&$\Lambda_b\left(\frac12^-\right)$\\
parent& $0.352\pm0.017$& $-10.83\pm0.65$& $0.376\pm0.014$&
$-12.82\pm0.58$\\
1 daughter &$0.397\pm0.015$&$-14.33\pm0.64$ & $0.419\pm0.010$&
$-16.33\pm0.45$\\
2 daughter &$0.438\pm0.015$&$-17.82\pm0.68$ & $0.460\pm0.008$&
$-19.84\pm0.36$\\
$b\{q,q\}$ &$\Sigma_b\left(\frac12^+\right)$&&$\Sigma_b^*\left(\frac32^+\right)$\\
parent& $0.368\pm0.014$& $-12.03\pm0.55$& $0.404\pm0.012$&
$-12.34\pm0.46$\\
1 daughter &$0.390\pm0.016$&$-14.59\pm0.67$ & $0.428\pm0.014$&
$-15.12\pm0.58$\\
2 daughter &$0.414$&$-17.42$ & $0.472$&
$-18.95$\\
$b[s,q]$ &$\Xi_b\left(\frac12^+\right)$&&$\Xi_b\left(\frac12^-\right)$\\
parent& $0.349\pm0.019$& $-11.49\pm0.80$& $0.381\pm0.014$&
$-13.88\pm0.60$\\
1 daughter &$0.399\pm0.016$&$-15.27\pm0.69$ & $0.423\pm0.011$&
$-17.40\pm0.49$\\
2 daughter &$0.440\pm0.015$&$-18.87\pm0.70$ & $0.465\pm0.008$&
$-21.03\pm0.40$\\
$b\{s,q\}$ &$\Xi_b'\left(\frac12^+\right)$&&$\Xi_b^*\left(\frac32^+\right)$\\
parent& $0.356\pm0.014$& $-12.16\pm0.58$& $0.386\pm0.014$&
$-12.33\pm0.57$\\
1 daughter &$0.360\pm0.053$&$-14.01\pm2.31$ & $0.386\pm0.061$&
$-14.11\pm2.62$\\
2 daughter &$0.346$&$-14.95$ & $0.364$&
$-14.83$\\
$b\{s,s\}$ &$\Omega_b\left(\frac12^+\right)$&&$\Omega_b^*\left(\frac32^+\right)$\\
parent& $0.365\pm0.013$& $-13.04\pm0.58$& $0.389\pm0.011$&
$-13.02\pm0.47$\\
1 daughter &$0.378\pm0.052$&$-15.30\pm2.35$ & $0.401\pm0.062$&
$-15.33\pm2.74$\\
2 daughter &$0.373$&$-16.79$ & $0.391$&
$-16.66$\\
\end{tabular}
 \end{ruledtabular}
\end{table}

\begin{table}
  \caption{Fitted parameters $\beta$, $\beta_0$ for the slope and intercept of the $(n_r,M^2)$  Regge
    trajectories for heavy baryons.} 
  \label{tab:rtn}
\begin{ruledtabular}
\begin{tabular}{cccccc}
& & \multicolumn{2}{c}{\hspace{-1.2cm}\underline{\hspace{2.4cm}$Q=c$\hspace{2.4cm}}}\hspace{-0.8cm}& \multicolumn{2}{c}{\underline{\hspace{2.2cm}$Q=b$\hspace{2.2cm}}}\\
Baryon& $Qd$ state &$\beta$ (GeV$^{-2}$)& $\beta_0$&$\beta$ (GeV$^{-2}$)& $\beta_0$\\
\hline
$\Lambda_Q\left(\frac12^+\right)$&$S$& $0.472\pm0.010$& $-2.543\pm0.099$& $0.238\pm0.011$&$-7.722\pm0.489$\\
$\Lambda_Q\left(\frac12^-\right)$&$P$& $0.494\pm0.006$&
$-3.363\pm0.059$& $0.248\pm0.010$&$-8.848\pm0.453$\\
$\Lambda_Q\left(\frac32^-\right)$&$P$& $0.495\pm0.005$&
$-3.444\pm0.053$& $0.249\pm0.010$&$-8.925\pm0.446$\\
$\Lambda_Q\left(\frac52^+\right)$&$D$& $0.508\pm0.003$&
$-4.225\pm0.030$& $0.260\pm0.009$&$-10.05\pm0.38$\\
$\Lambda_Q\left(\frac72^-\right)$&$F$& $0.508\pm0.005$&
$-4.824\pm0.059$& $0.280\pm0.008$&$-11.55\pm0.36$\\
$\Sigma_Q\left(\frac12^+\right)$&$S$& $0.445\pm0.009$& $-2.696\pm0.089$& $0.233\pm0.009$&$-7.942\pm0.366$\\
$\Sigma_Q\left(\frac12^-\right)$&$P$& $0.469\pm0.006$&
$-3.694\pm0.070$& $0.246\pm0.005$&$-9.190\pm0.238$\\
$\Sigma_Q\left(\frac52^-\right)$&$P$& $0.472\pm0.006$&
$-3.693\pm0.069$& $0.249\pm0.005$&$-9.256\pm0.234$\\
$\Sigma_Q\left(\frac12^+\right)$&$D$& $0.474$&
$-4.384$& $0.238$&$-9.466$\\
$\Xi_Q\left(\frac12^+\right)$&$S$& $0.444\pm0.010$& $-2.805\pm0.118$& $0.234\pm0.010$&$-8.064\pm0.464$\\
$\Xi_Q\left(\frac12^-\right)$&$P$& $0.465\pm0.007$&
$-3.658\pm0.078$& $0.251\pm0.010$&$-9.527\pm0.461$\\
$\Xi_Q\left(\frac32^-\right)$&$P$& $0.465\pm0.006$&
$-3.729\pm0.073$& $0.252\pm0.010$&$-9.589\pm0.452$\\
$\Xi_Q\left(\frac52^+\right)$&$D$& $0.479\pm0.004$&
$-4.540\pm0.049$& $0.263\pm0.009$&$-10.72\pm0.40$\\
$\Xi_Q\left(\frac72^-\right)$&$F$& $0.488\pm0.001$&
$-5.301\pm0.017$& $0.282\pm0.008$&$-12.28\pm0.37$\\
$\Xi'_Q\left(\frac12^+\right)$&$S$& $0.432\pm0.006$& $-2.871\pm0.060$& $0.233\pm0.008$&$-8.279\pm0.369$\\
$\Xi'_Q\left(\frac12^-\right)$&$P$& $0.448\pm0.007$&
$-3.880\pm0.087$& $0.237\pm0.011$&$-9.276\pm0.490$\\
$\Xi'_Q\left(\frac52^-\right)$&$P$& $0.450\pm0.007$&
$-3.883\pm0.078$& $0.240\pm0.010$&$-9.379\pm0.459$\\
$\Xi'_Q\left(\frac12^+\right)$&$D$& $0.451$&
$-4.541$& $0.236$&$-9.829$\\
$\Omega_Q\left(\frac12^+\right)$&$S$& $0.414\pm0.006$& $-3.004\pm0.069$& $0.232\pm0.008$&$-8.609\pm0.385$\\
$\Omega_Q\left(\frac12^-\right)$&$P$& $0.429\pm0.008$&
$-4.032\pm0.098$& $0.237\pm0.011$&$-9.608\pm0.498$\\
$\Omega_Q\left(\frac52^-\right)$&$P$& $0.432\pm0.007$&
$-4.049\pm0.088$& $0.240\pm0.010$&$-9.701\pm0.488$\\
$\Omega_Q\left(\frac12^+\right)$&$D$& $0.431$&
$-4.654$& $0.235$&$-10.07$\\
\end{tabular}
 \end{ruledtabular}
\end{table}

The obtained results allow us to determine the possible quantum numbers of the
observed heavy baryons and prescribe them to a particular Regge trajectory. 
In the ($J, M^2$) plane there are three trajectories for which three
experimental candidates are available (parent trajectories for the
$\Lambda_c\left(\frac12^+\right)$ in Fig.~\ref{fig:lambda_j}a, for the
$\Xi_c\left(\frac12^+\right)$  in Fig.~\ref{fig:xi_sd_j}a and for the
$\Xi_c^*\left(\frac32^+\right)$ in Fig.~\ref{fig:xi_vd_j}b) and two
trajectories with two experimental candidates
(parent trajectories for the $\Sigma_c\left(\frac12^+\right)$ in
Fig.~\ref{fig:sigma_j}a and for the
$\Xi_c\left(\frac12^-\right)$ in Fig.~\ref{fig:xi_sd_j}b). On the other
hand, in the ($n_r, M^2$) plane there are three trajectories with two
experimental candidates (the $\Lambda_c\left(\frac12^+\right)$ and the
$\Lambda_c\left(\frac12^-\right)$  in Fig.~\ref{fig:lambda_n} and the
$\Xi_c\left(\frac12^+\right)$ in Fig.~\ref{fig:xi_vd_n}). All
experimental points fit well to the corresponding  Regge
trajectories obtained in our model. 

From Tables~\ref{tab:lq}, \ref{tab:sq} and
Figs.~\ref{fig:lambda_j}, \ref{fig:sigma_j}, \ref{fig:lambda_n}, \ref{fig:sigma_n} we see that the
$\Lambda_c(2765)$ (or $\Sigma_c(2765)$),~\footnote{It is important to note that
the $J^P$ quantum numbers for most excited heavy baryons have not
been determined experimentally, but are assigned by PDG on the basis of
quark model predictions. For some excited charm
baryons such as the $\Lambda_c(2765)$, $\Lambda_c(2880)$ and
$\Lambda_c(2940)$  it is even not known if they are excitations of the
$\Lambda_c$ or $\Sigma_c$.} if it is indeed the $\Lambda_c$
state,  can be interpreted in our model as the first radial ($2S$)
excitation of the $\Lambda_c$. If instead it is the $\Sigma_c$ state, then it can
be identified as its first orbital excitation ($1P$) with
$J=\frac32^-$ (see Table~\ref{tab:sq}). The $\Lambda_c(2880)$ baryon
corresponds to the second orbital excitation ($2D$) with $J=\frac52^+$,
fitting nicely the parent $\Lambda_c$ Regge trajectory in the ($J,M^2$)
plane (see Fig.~\ref{fig:lambda_j}a). Such prescription is in accord
with the experimental evidence coming from the $\Sigma_c(2455) \pi$
decay angular distribution \cite{pdg}. The other charmed baryon, denoted as
$\Lambda_c(2940)$, probably has $I=0$, since it was discovered in the $pD^0$ mass
spectrum and not observed in $p D^+$ channel, but $I=1$ is not ruled out \cite{pdg}. If
it is really the $\Lambda_c$, state then it could be both an orbitally and
radially excited ($2P$) state with $J=\frac12^-$, whose mass is
predicted to be about 40 MeV heavier (see Fig.~\ref{fig:lambda_j}b). A better agreement with
experiment (within few MeV) is achieved, if the $\Lambda_c(2940)$ is
interpreted as the first radial excitation ($2S$) of the $\Sigma_c$
with $J=\frac32^+$ (see Fig.~\ref{fig:sigma_j}b). The $\Sigma_c(2800)$ can be identified with one of
the first orbital ($1P$) excitations of the $\Sigma_c$  with
$J=\frac12^-$ or $\frac32^- $ which have very close masses compatible
with experimental value within errors (see Table~\ref{tab:sq}).  

The results for masses and the Regge trajectories of the $\Xi_Q$ baryons
both with the scalar and axial vector diquarks are given in
Tables~\ref{tab:xs}, \ref{tab:xv} and 
Figs.~\ref{fig:xi_sd_j}, \ref{fig:xi_vd_j}, \ref{fig:xi_sd_n},
\ref{fig:xi_vd_n}. From these tables and plots we see that the
$\Xi_c(2790)$ and $\Xi_c(2815)$ can be assigned to the first orbital
($1P$) excitations of the $\Xi_c$ containing a scalar diquark with $J=\frac12^-$
and  $J=\frac32^-$, respectively. On the other hand, the charmed
baryon $\Xi_c(2930)$ can be considered as either the $J=\frac12^-$,
$J=\frac32^-$  or $J=\frac52^-$ state (all these states are predicted to have close masses)
corresponding to  the first orbital ($1P$) excitations of the $\Xi_c'$
with an axial vector diquark.  While the $\Xi_c(2980)$ can be viewed as  the
first radial ($2S$) excitation with $J=\frac12^+$ of the $\Xi_c'$, the
$\Xi_c(3055)$ and  $\Xi_c(3080)$ baryons can be interpreted  as a second
orbital ($2D$) excitations of the $\Xi_c$ containing a scalar diquark
with $J=\frac32^+$ and  $J=\frac52^+$, and the $\Xi_c(3123)$ can be
viewed as the corresponding ($2D$) excitation of the $\Xi_c'$  with
$J=\frac72^+$. 

For the $\Omega_c$ baryons as well as for all bottom baryons only masses
of ground states are known \cite{pdg}, most of which were measured
recently. Our original predictions for the ground states \cite{hbar}
of these baryons are very close to the values found in the present
analysis (see Tables~\ref{tab:lq}-\ref{tab:om}) and agree well with
measurements \cite{pdg}.  

The detailed comparison of our predictions for the masses
of the ground and lowest excited states of heavy baryons with the results of
other theoretical calculations \cite{ci,mmmp,gvv} is given in Table~8 of
Ref.~\cite{excbar}.   

\subsection{Relations between parameters of the Regge trajectories}
\label{sec:rp}

The slopes of the Regge trajectories, given in Tables~\ref{tab:rtj}, \ref{tab:rtn},
follow in both planes the pattern previously observed for light and heavy
mesons \cite{lregge,hlmr}. They decrease with the increase of the
diquark mass or with 
the increase of the heavy quark mass. The latter decrease is even
more pronounced. The mass dependence of the parameters of the Regge
trajectories is the result of the flavour dependence of the potential
(\ref{eq:v}). Such behaviour agrees with the phenomenological
consideration of Ref.~\cite{gww}.  

It was argued in the literature on the basis of different models within
QCD (see e.g. \cite{k,bg,gww} and references therein) that  the
parameters of the Regge trajectories for the baryon 
multiplets with given $J^P$ and  different quark constituents can be related
by a set of relations, which for heavy baryons is given by:

(a) the additivity of inverse slopes
\begin{equation}
  \label{eq:aisl}
  \frac1{\alpha(\Sigma_Q)}+\frac1{\alpha(\Omega_Q)}=\frac2{\alpha(\Xi'_Q)},
\end{equation}

(b) the additivity of intercepts
\begin{equation}
  \label{eq:aint}
  \alpha_0(\Sigma_Q)+\alpha_0(\Omega_Q)=2\alpha_0(\Xi'_Q),
\end{equation}

(c) the factorization of slopes
\begin{equation}
  \label{eq:fsl}
  \alpha(\Sigma_Q)\alpha(\Omega_Q)=\alpha^2(\Xi'_Q).
\end{equation}
Such relations were extensively used in the literature for obtaining
different linear and quadratic mass relations for light and heavy
baryons (see e.g. \cite{gww} and references therein) and for obtaining
on their basis predictions for the baryon masses.  
However, it was argued in Ref.~\cite{bg} that relations (\ref{eq:aisl})
and (\ref{eq:fsl}) are incompatible for heavy baryons. Moreover, it was
shown there that the factorization of slopes
(\ref{eq:fsl}) violates the heavy quark limit for heavy baryons, but
this violation introduces rather small errors (less than 15\%).   
The test of the validity of these relations in our model is given
in Table~\ref{tab:slr}. It is not surprising that all relations for
the slopes are satisfied within the error bars both for parent and
daughter trajectories, since the slopes have close values. Let us
mention that the slopes of the parent Regge trajectories in the ($J,M^2$)
plane, obtained in our approach have close values to the ones found in the
phenomenological analysis \cite{gww}, based on the different mass
relations for light and heavy baryons.        

\begin{table}
  \caption{Test of the relations between parameters of the
    heavy-baryon Regge trajectories in the ($J,M^2$) plane.} 
  \label{tab:slr}
\begin{ruledtabular}
\begin{tabular}{cc@{$\!$}cc@{$\!$}c@{$\!$}ccc}
$J^P$& Traject.& $
\frac1{\alpha(\Sigma_Q)}+\frac1{\alpha(\Omega_Q)}$ &
$\frac2{\alpha(\Xi'_Q)}$ & $\alpha_0(\Sigma_Q)+\alpha_0(\Omega_Q)$ &
$2\alpha_0(\Xi'_Q)$ &
$\alpha(\Sigma_Q)\alpha(\Omega_Q)$ & $\alpha^2(\Xi'_Q)$ \\
& & (GeV$^2$) & (GeV$^2$)& & &(GeV$^{-4}$)&(GeV$^{-4}$)\\
\hline
&$Q=c$\\
$\frac12^+$& parent& $3.10\pm 0.13$ & $3.11\pm 0.15$& $-7.73\pm 0.30$
& $-7.71\pm 0.42$ & $0.418\pm 0.035$& $0.414\pm 0.041$ \\
$\frac12^+$&1 daughter& $3.23\pm 0.12$ & $3.32\pm 0.14$& $-10.20\pm 0.47$
& $-9.78\pm 0.54$ & $0.388\pm 0.028$& $0.364\pm 0.031$ \\
$\frac12^+$&2 daughter& $3.25$ & $3.30$& $-13.16$
& $-12.83$ & $0.384$& $0.367$ \\
$\frac32^+$& parent& $2.74\pm 0.07$ & $2.76\pm 0.07$& $-7.36\pm 0.37$
& $-7.33\pm 0.38$ & $0.537\pm 0.029$& $0.527\pm 0.028$ \\
$\frac32^+$&1 daughter& $2.92\pm 0.04$ & $2.99\pm 0.03$& $-9.70\pm 0.15$
& $-9.23\pm 0.10$ & $0.477\pm 0.010$& $0.445\pm 0.009$ \\
$\frac32^+$&2 daughter& $2.73$ & $2.82$& $-14.03$
& $-13.73$ & $0.542$& $0.501$ \\
&$Q=b$\\
$\frac12^+$& parent& $5.46\pm 0.20$ & $5.62\pm 0.22$& $-25.1\pm 1.1$
& $-24.3\pm 1.2$ & $0.134\pm 0.010$& $0.127\pm 0.010$ \\
$\frac12^+$&1 daughter& $5.26\pm 0.48$ & $5.67\pm 0.83$& $-29.9\pm 3.0$
& $-28.0\pm 4.6$ & $0.148\pm 0.026$& $0.132\pm 0.038$ \\
$\frac12^+$&2 daughter& $5.10$ & $5.78$& $-34.2$
& $-29.9$ & $0.154$& $0.120$ \\
$\frac32^+$& parent& $5.05\pm 0.15$ & $5.19\pm 0.19$& $-25.4\pm 0.9$
& $-24.7\pm 1.1$ & $0.157\pm 0.009$& $0.149\pm 0.011$ \\
$\frac32^+$&1 daughter& $4.89\pm 0.47$ & $5.31\pm 0.84$& $-30.4\pm 3.3$
& $-28.2\pm 5.2$ & $0.172\pm 0.032$& $0.153\pm 0.047$ \\
$\frac32^+$&2 daughter& $4.68$ & $5.49$& $-35.6$
& $-29.7$ & $0.184$& $0.133$ \\
\end{tabular}
 \end{ruledtabular}
\end{table}
It is important to compare the values of the slopes of the Regge trajectories for
heavy baryons, heavy-light and light mesons.
From the comparison of the heavy baryon slopes in Tables~\ref{tab:rtj},
\ref{tab:rtn} we see that the $\alpha$ values are systematically larger than
the $\beta$ ones. The ratio of their mean values is about 1.5 both for
the charmed and bottom baryons. This value of the ratio is very close
to the one found for the heavy-light mesons \cite{hlmr} and is slightly
larger than the one (1.3) obtained for the light mesons  \cite{lregge}. 

 From comparison of
Tables~\ref{tab:rtj}, \ref{tab:rtn} and Tables~4, 5 of 
Ref.~\cite{hlmr} we find that for the same flavour of the heavy quark
the heavy baryon slopes have higher values than the heavy-light meson
ones. It is interesting that the ratios of the heavy baryon to
heavy-light meson slopes ($\alpha_{Qqq}/\alpha_{Q\bar q}$ and
$\beta_{Qqq}/\beta_{Q\bar q}$)  have very close values, which are about
1.4, both in the ($J, M^2$) and in ($n_r, M^2$) planes. Note that
light baryons and light mesons have almost equal values of the Regge
trajectory slopes (see Ref.~\cite{kr} and references therein).      

\section{Conclusions}
\label{sec:conc}

In this paper the spectroscopy of charmed and bottom baryons was studied in the
framework of the quark-diquark picture in the relativistic quark
model. The heavy baryon was considered as a heavy-quark--light-diquark
bound system in which excitations occur only between a heavy quark and
a light diquark. The light diquarks were considered only in the ground
(either scalar or  axial vector) state. The diquarks were not
treated as point-like objects. Their internal structure was taken into
account by including form factors of the diquark-gluon interaction in
terms of the diquark wave functions. The dynamics of light quarks
inside a diquark as well as the dynamics of a light diquark and a heavy
quark inside a baryon were treated completely relativistically without
application of either the nonrelativistic $v/c$ or heavy quark $1/m_Q$
expansions. Such nonperturbative approach is especially important for
the highly excited charmed baryon states, where the heavy quark
expansion is not adequate enough. It is important to emphasize that
all parameters of our relativistic quark model such as quark masses
and parameters of the interquark potential were fixed previously form
the investigation of meson mass spectra and decay processes. Thus our
model provides a unified universal description of meson and baryon
properties.    

We calculated the masses of ground, orbitally and radially excited
heavy baryons up to rather high excitations ($L=5$ and
$n_r=5$). This allowed us to
construct the Regge trajectories both in the $(J,M^2)$ and  $(n_r,M^2)$
planes. It was found that they are almost linear, parallel and
equidistant. The available experimental data nicely fit to
them. The assignment of the experimentally observed heavy baryons to
the particular Regge trajectories was carried out. This allowed us to
determine the quantum numbers of the excited heavy baryons. It was
found that all currently available experimental data can be well
described in the relativistic quark-diquark picture, which predicts
significantly less states than the genuine three-body
picture. 

The comparison of the slopes of the Regge trajectories of
heavy baryons and heavy-light mesons was given. It was found that
the slope values of heavy baryons are approximately 1.4 times higher
than the ones of heavy mesons with the same flavour of the heavy quark.  

%\vspace*{0.5cm}
\acknowledgements
The authors are grateful to M. M\"uller-Preussker  for support  and to
V. Matveev, V. Savrin and M. Wagner for discussions.  Two of us
(R.N.F. and V.O.G.)  acknowledge  the support by the {\it Deutsche
Forschungsgemeinschaft} under contract Eb 139/6-1.

\end{document}